\documentclass[10pt,two column,amsmath,amssymb,aps]{revtex4}
\usepackage{graphicx}
\newcommand{\be}{\begin{equation}}
\newcommand{\bea}{\begin{eqnarray}}
\newcommand{\ee}{\end{equation}}
\newcommand{\eea}{\end{eqnarray}}

\usepackage{amsmath}
\usepackage{bbold}
 \usepackage{float} 
 
\begin{document}
\title{Efficient excitation of nonlinear phonons via chirped mid-infrared pulses: induced structural phase transitions}
\author{A.P. Itin$^{1,2}$ and  M.I. Katsnelson$^1$}
\affiliation{$^1$Radboud University, Nijmegen, The Netherlands, \\$^2$Space Research Institute, Moscow, Russia}
\begin {abstract}
Nonlinear phononics  play  important role in strong laser-solid interactions. We discuss a dynamical protocol for efficient phonon excitation, considering recent inspiring proposals:  inducing ferroelectricity in paraelectric material such as KTaO$_3$, and inducing structural deformations in cuprates (e.g  La$_2$CuO$_4$)  [A. Subedi et.al, Phys.Rev. B 89, 220301 (2014), Phys.Rev. B 95, 134113 (2017)]. High-frequency phonon modes  are driven by mid-infrared pulses, and coupled to lower-frequency modes those indirect excitation causes structural deformations.   We study in a more detail the case of KTaO$_3$ without strain, where (at first glance) it was not possible to excite the needed low frequency phonon mode  by resonant driving of the higher frequency one.  Behaviour of the system is explained using a reduced model of coupled driven nonlinear oscillators.  We find a dynamical mechanism which prevents effective excitation at resonance driving.  In order to  induce ferroelectricity  we employ driving with sweeping frequency, realizing so called capture into resonance.  The method works for realistic femtosecond pulses and can be applied to many other related systems.
 \end{abstract}
 \maketitle
 
Research in ultrafast light-control of materials have attracted a lot of interest recently \cite{Phononics, SubediFerro, Tokura, Hauser, Cavalleri07,Cavalleri08,Cavalleri11,Cavalleri12,Cavalleri15}.  Intense mid-infrared pulses have been used to directly control the dynamical degrees of freedom of the crystal lattice \cite{Cavalleri07,Cavalleri08,Cavalleri11,Cavalleri12,Cavalleri15}, in particular, inducing melting of orbital and magnetic  orders. In many recent suggestions and experiments phonon modes are driven indirectly:  a laser  drives a high-frequency infrared-active phonon mode which then excites required modes not easily accessible by direct drive  by means of nonlinear couplings. Here we suggest to add a new useful tool to the arsenal of nonlinear phononics: capture into a resonance. Such phenomenon is encountered in classical and celestial mechanics \cite{Sinclair, Greenberg,AKN} and was employed, e.g. in plasma and accelerator physics \cite{Veksler,McMillan,Neishtadt,surfatron,surfatron2}.  In a nonlinear system with near-resonant driving, as amplitude of perturbation grows, frequency of the system varies, and it stops to absorb energy efficiently. Driving with changing frequency enables to  sustain resonance relation between the drive and the system \cite{ANeishtadt}. 

 \begin{figure}
\includegraphics[width=41mm]{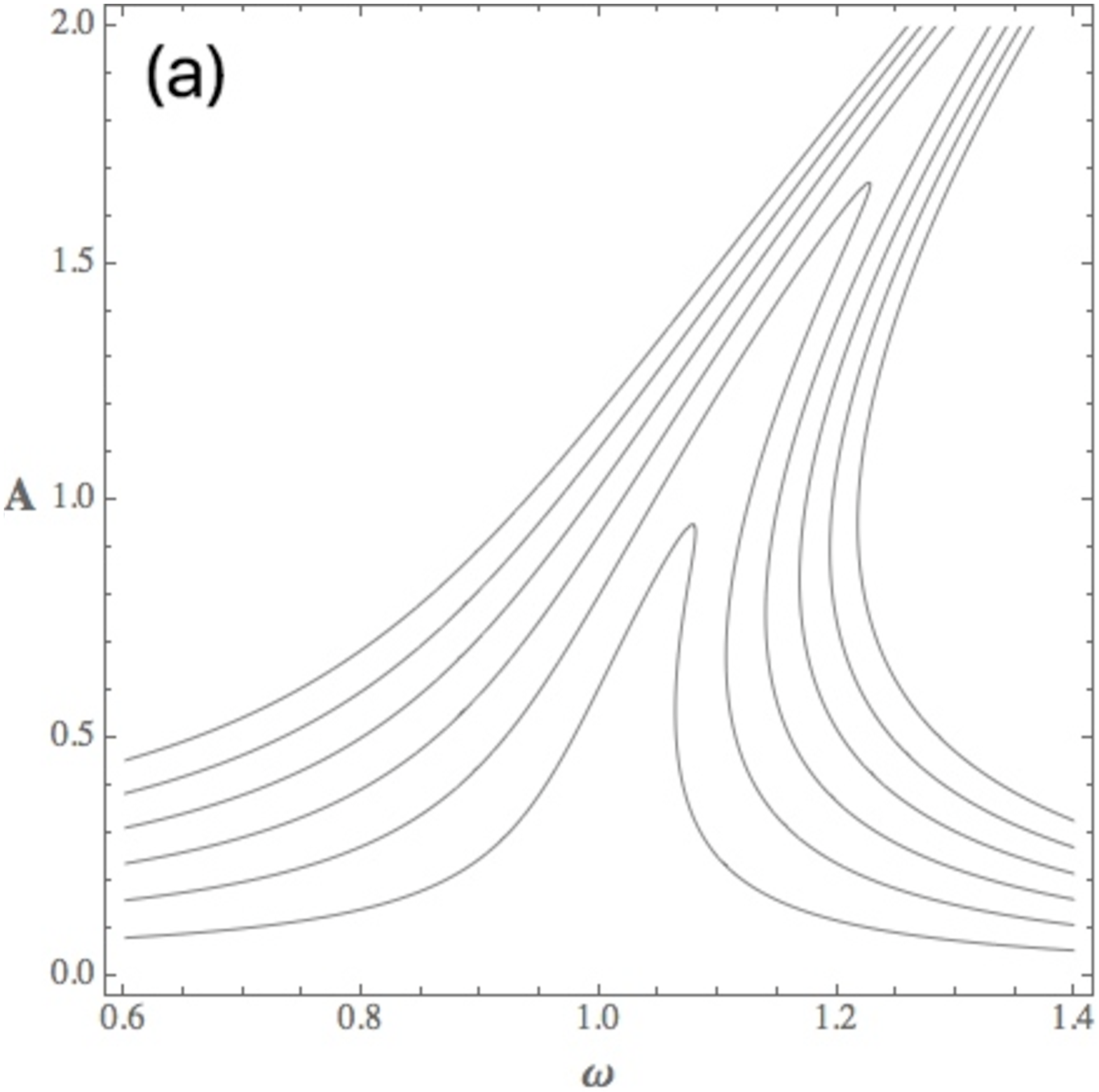}
\includegraphics[width=41mm]{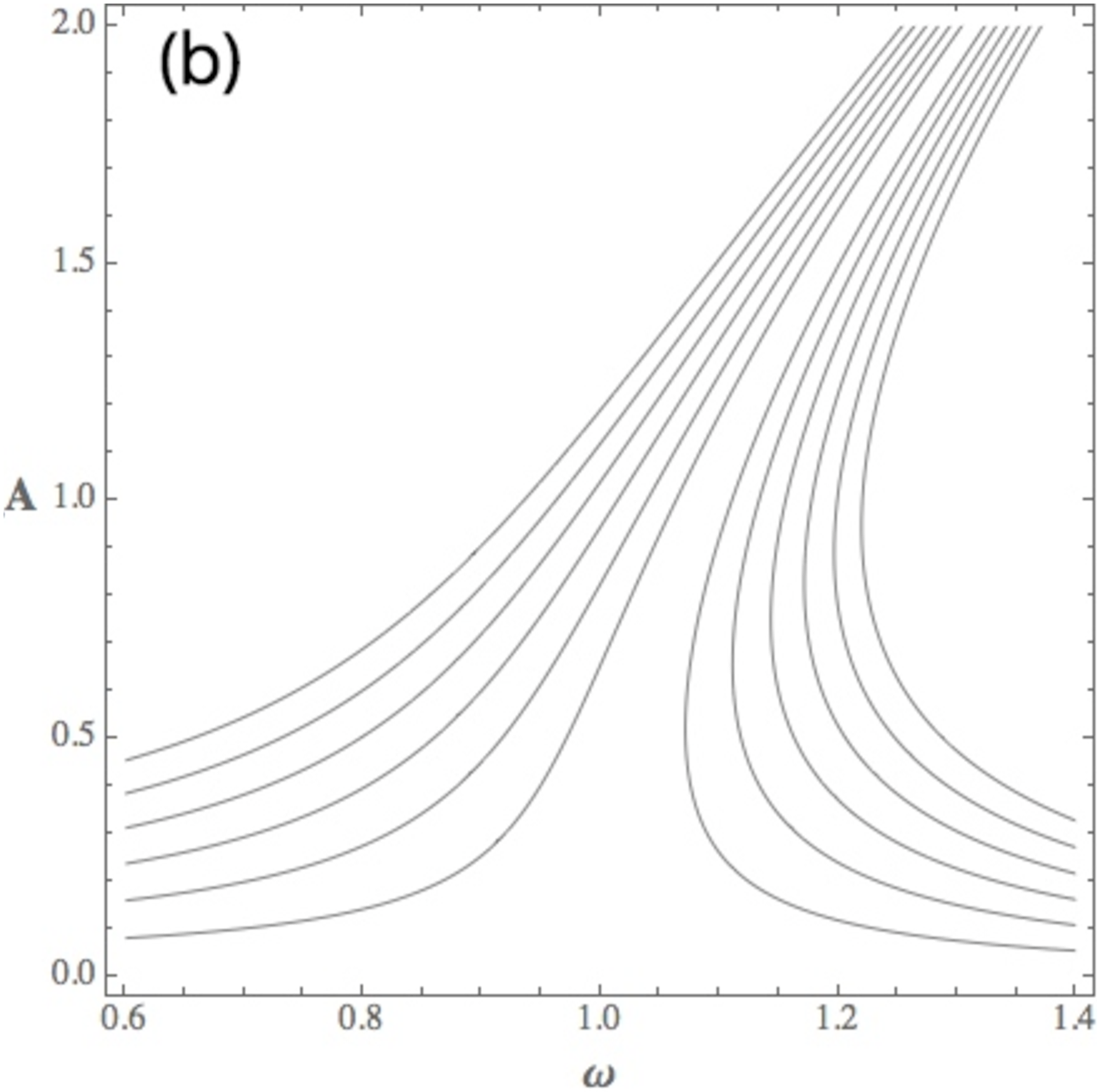}
\includegraphics[width=41mm]{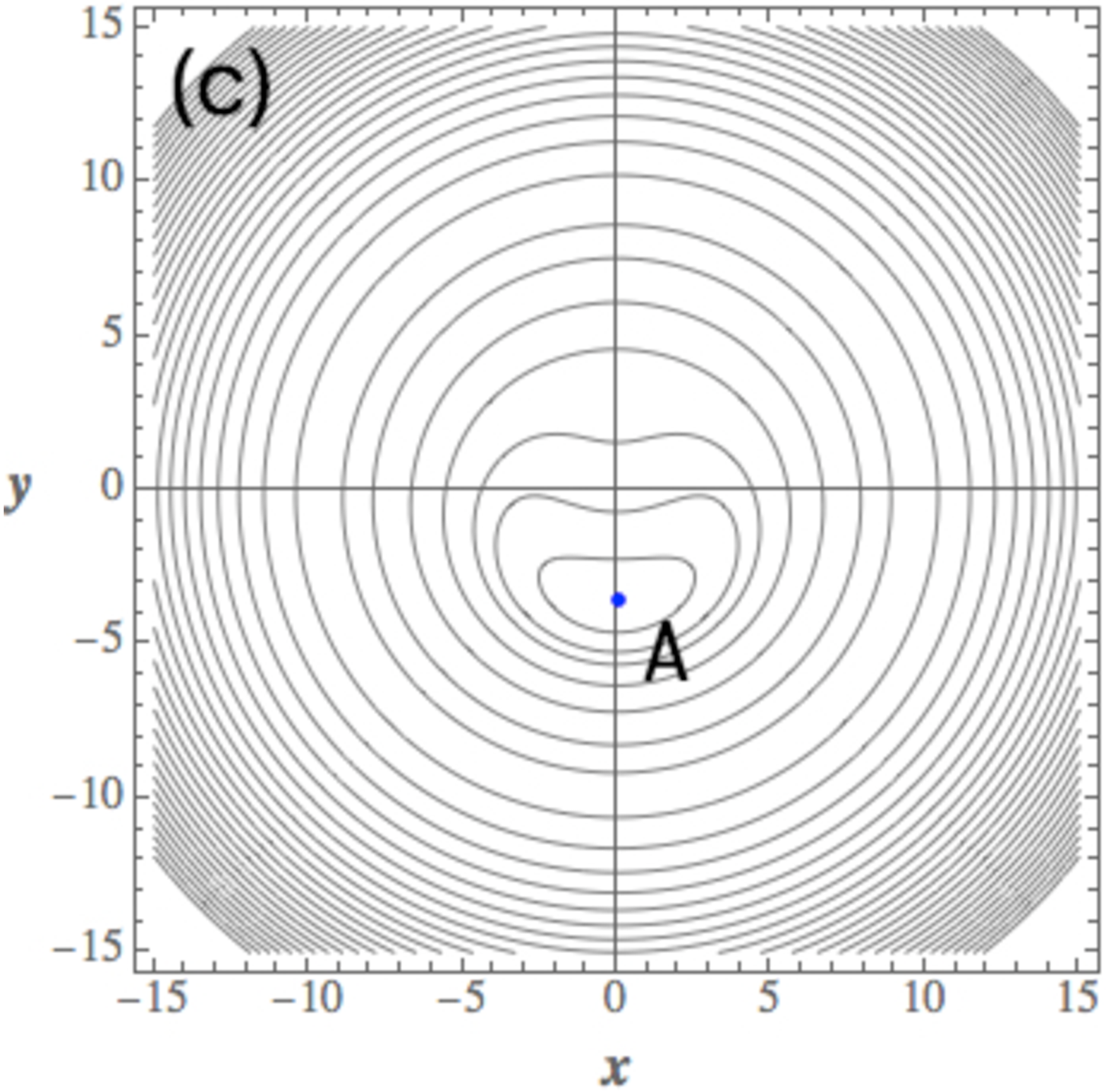}
\includegraphics[width=41mm]{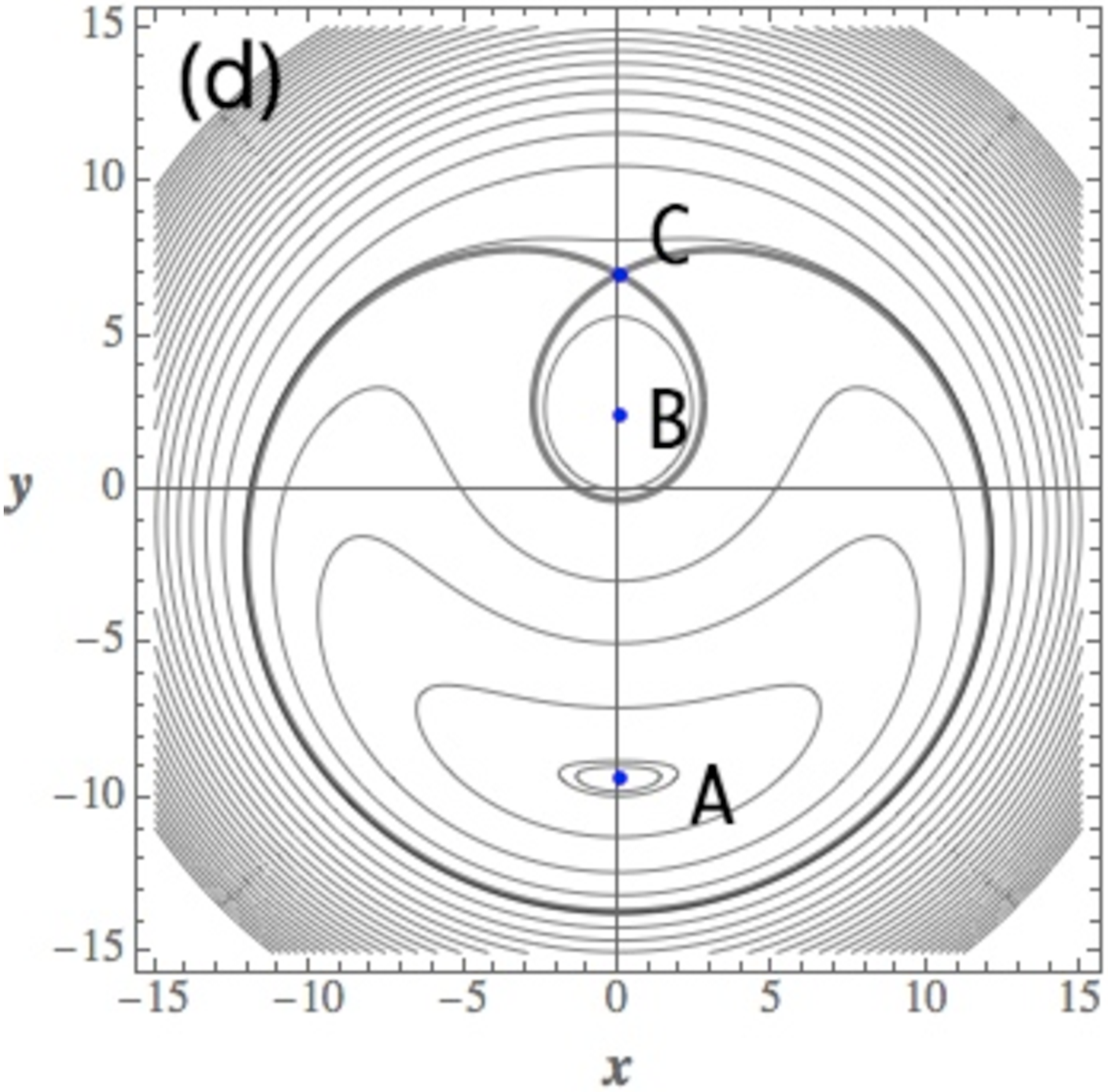}
\caption{Top panel: steady states of the driven Duffing oscillator (\ref{ duff})  as a function of $\omega \equiv \Omega/\Omega_{0}$ at several values of the driving force amplitude $F_0$ ($A$ is the  amplitude of  $Q$). (a) damped Duffing oscillator (b) undamped  Duffing oscillator. Curves from bottom to top correspond to: $E_0=5,10,15,20,25,30$ MeV  cm$^{-1}$.   Bottom panel: phase portraits   of the effective Hamiltonian (\ref{Hres})  (c).  $\lambda< \lambda_* = \frac{3}{2} \mu^{2/3} $   (d)  $\lambda> \lambda_*  =  \frac{3}{2} \mu^{2/3}$.   \label{duffFig}}
\end{figure}
We find that in perovskite paraelectric KTaO$_3$ (being described below), as well as in other systems (like La$_2$CuO$_4$ cuprate (LCO)), it is possible to considerably excite a high-frequency phonon mode using a protocol based on capture into resonance, which requires a pulse with chirped frequency. This excitation makes a coupled low-frequency phonon mode dynamically unstable.
Pulses with frequency chirps on picosecond timescale have been generated e.g. in FELIX  \cite{FELIX}.

In all considered examples below, a phonon mode with (at least) quartic nonlinearity  is driven by a laser pulse,  creating effective potential for a  coupled lower-frequency phonon mode those excitation triggers structural phase transition.   Neglecting dissipation,  our starting Hamiltonian $ H = \frac{P^2}{2} +  \frac{\Omega_0^2 Q^2}{2} + c_4 Q^4 +c_6..  - Q F_0 \sin{\Phi (t)}, $
where $\Phi(t)$ is the phase of the driving  $\dot \Phi(t) = \Omega(t)$,  $\Omega_0$ is linear frequency of the driven phonon mode, $c_4, c_6..$ are anharmonic  coefficients, $F_0(t)$ is the amplitude of the external field.
Introducing symplectic coordinates $P =   \sqrt{ 2I  \Omega_0  } \cos \phi, $ $Q = \sqrt{ \frac{2I}{\Omega_0} } \sin \phi $
we then make a transformation to the resonance phase $\gamma = \phi - \Phi$ using the time-dependent generating function $ W =  J( \phi - \Phi)$.
The new Hamiltonian $H' = H - J \Omega$  can be averaged over the fast phase, with  the result (neglecting nonlinearities higher than the quartic for a while)
$ {\cal H} = \delta \Omega J + \frac{3}{2} c_4 \frac{J^2}{\Omega_0^2}  - \frac{1}{2} F_0 \sqrt{ 2J/\Omega_0}  \cos \gamma. $
Introducing now $x=  \sqrt{ 2J } \sin \gamma, $ $y=  \sqrt{ 2J } \cos \gamma, $ we get an effective Hamiltonian
 $H= \frac{3}{8}  \frac{c_4}{\Omega_0^2}  (x^2 +y^2)^2  +  \frac{\delta \Omega}{2} (x^2+y^2) -  \frac{F_0}{2 \sqrt{\Omega_0} y} .$
 Upon rescaling $H \to H/  \frac{3}{8}  \frac{c_4}{\Omega_0^2} , $ $t \to  t \frac{3}{8}  \frac{c_4}{\Omega_0^2}  $ and introduction of 
$\lambda =  -  \frac{\delta \Omega }{2}  / \frac{3}{8}  \frac{c_4}{\Omega_0^2}   $, $\mu = -  \frac{F_0}{2 \sqrt{\Omega_0} } $
we bring the Hamiltonian to the form
\be
H = (x^2 + y^2)^2 - \lambda(t) (x^2+ y^2) + \mu(t) y   \label{Hres}.
\ee
\begin{figure}
\includegraphics[width=42mm]{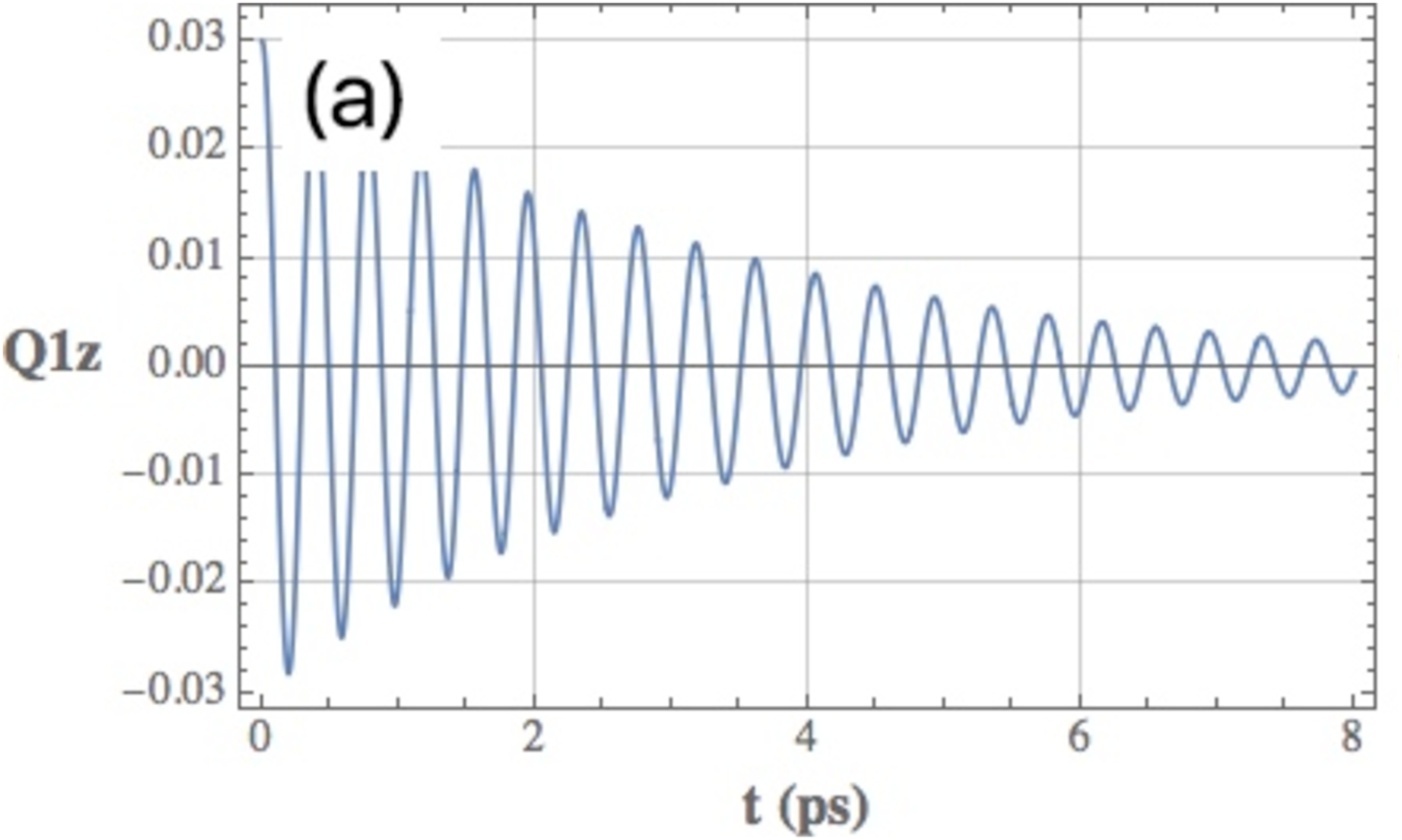}
\includegraphics[width=42mm]{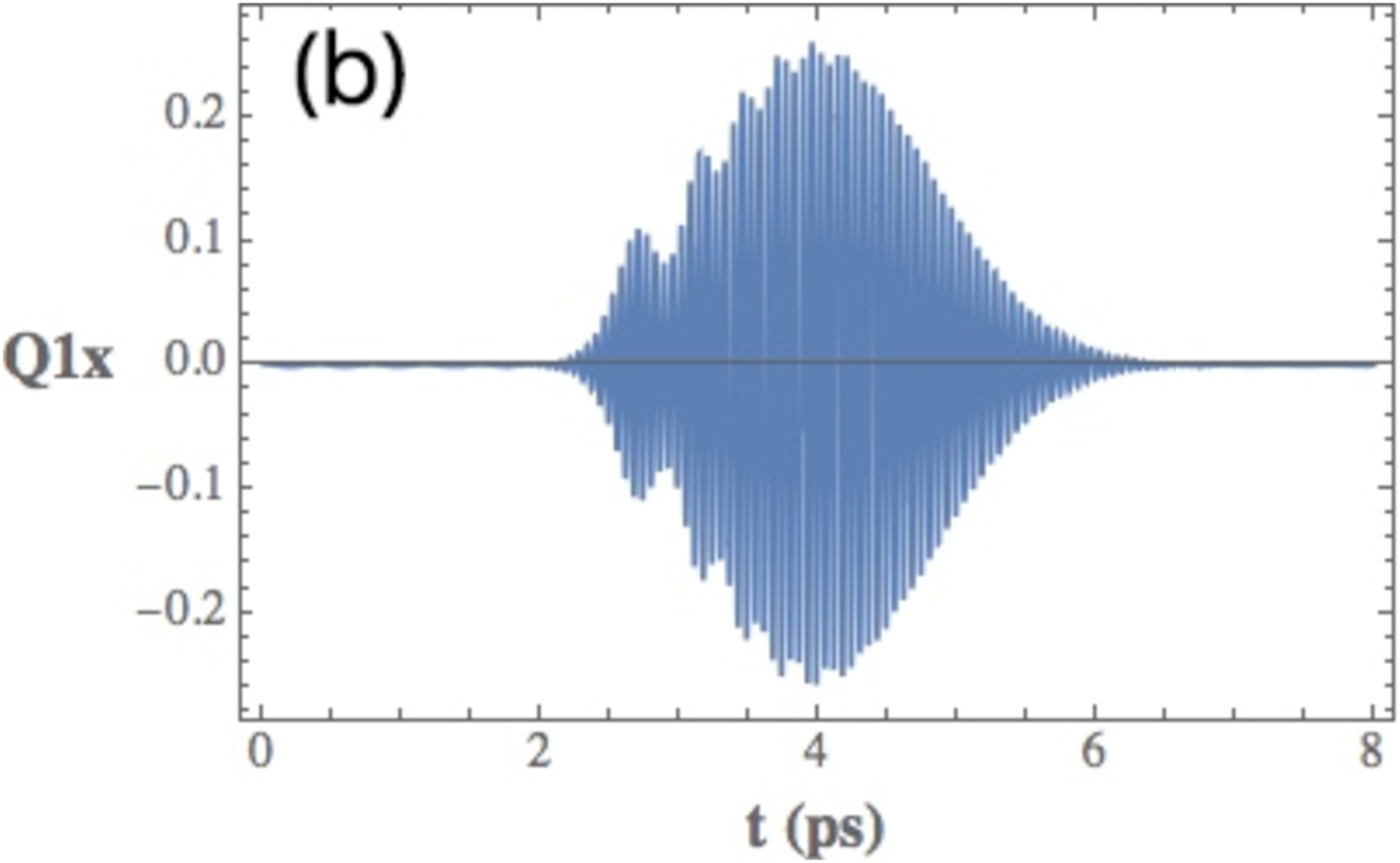}
\includegraphics[width=42mm]{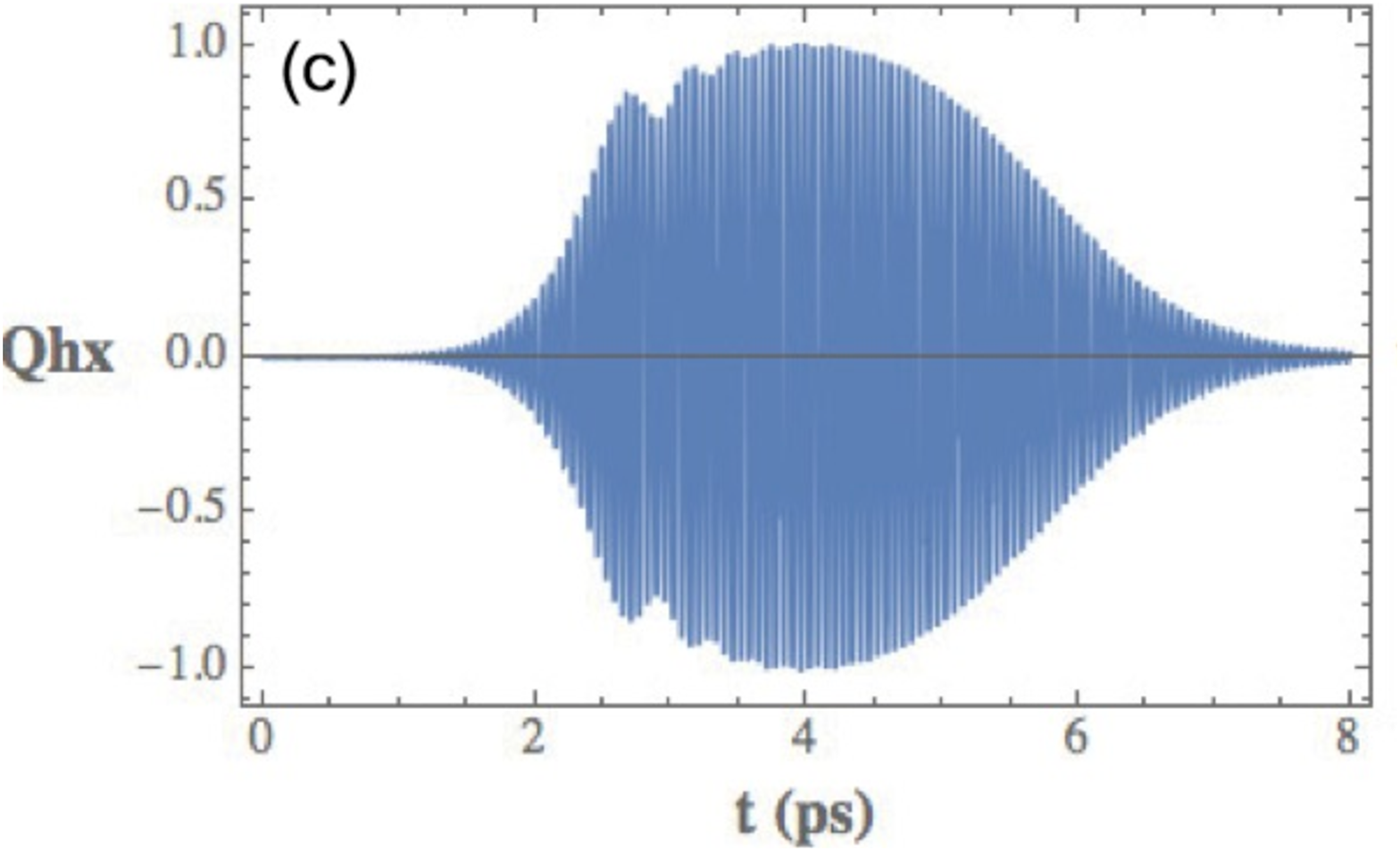}
\includegraphics[width=42mm]{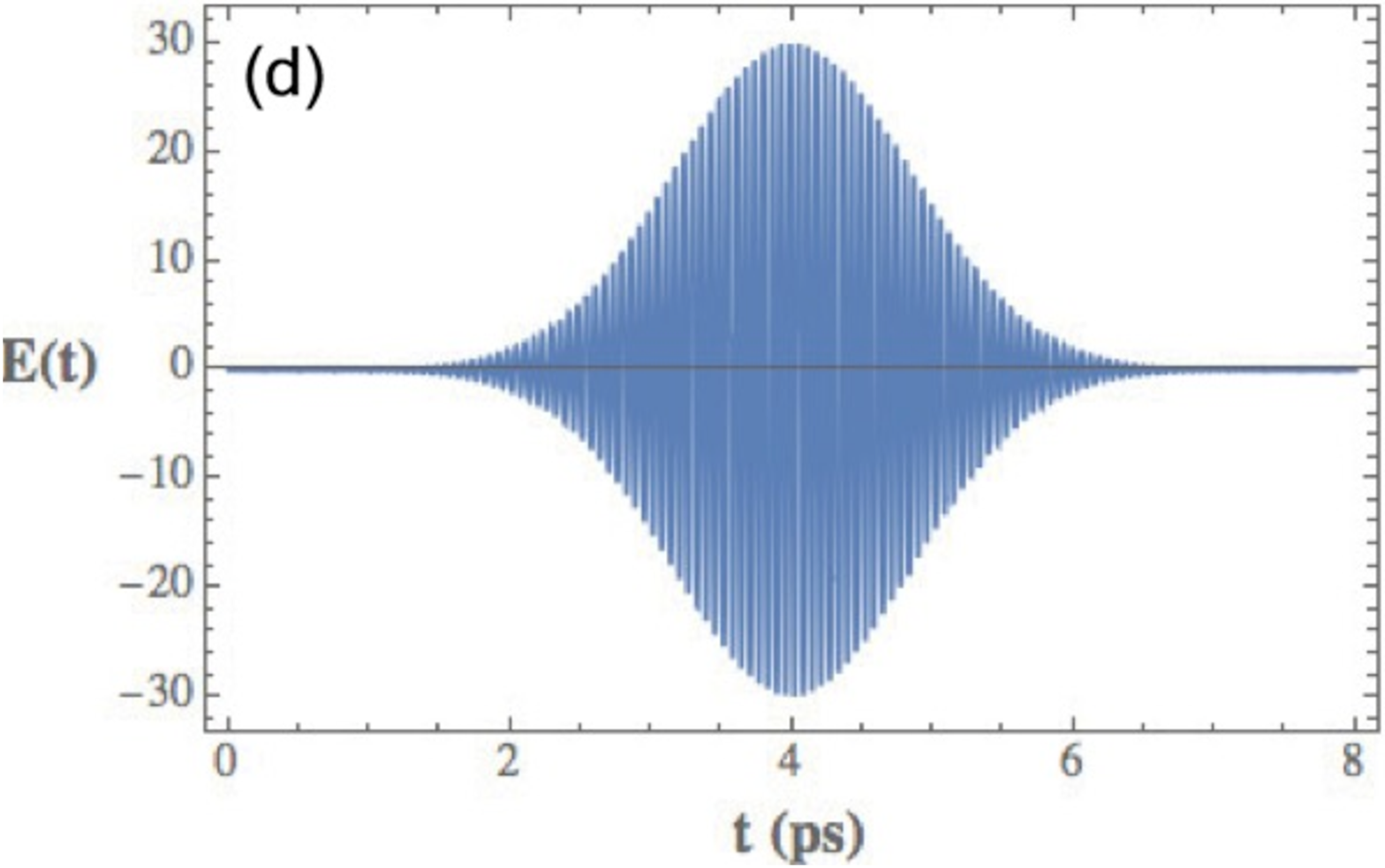}
\caption{ Dynamics of the system under resonant driving. (a-d):  $Q_{1z}, Q_{1x},Q_{hx}$ modes (in units of $\AA \sqrt{amu}$ \cite{SubediFerro}), and $E(t)$ pulse, correspondingly. (in units of MV cm$^{-1}$). Time is in picoseconds.  Parameters are: $E_0=30$ MV cm$^{-1}$, damping coefficients $\gamma_h,\gamma_1$ are $4\%$ of $\Omega_h, \Omega_1$, correspondingly. The width of the pulse is  $\sigma=$2ps. Similar to \cite{SubediFerro}, $Q_{1z}$ was given a small initial excitation to initiate its dynamics (simulating temperature fluctuations). Strong driving of $Q_{hx}$ and corresponding excitation of $Q_{1x}$ modes almost do not affect decay of the transverse mode $Q_{1z}$, and leaves it unexcited.   \label{E30} }
\end{figure}
\begin{figure}
\includegraphics[width=42mm]{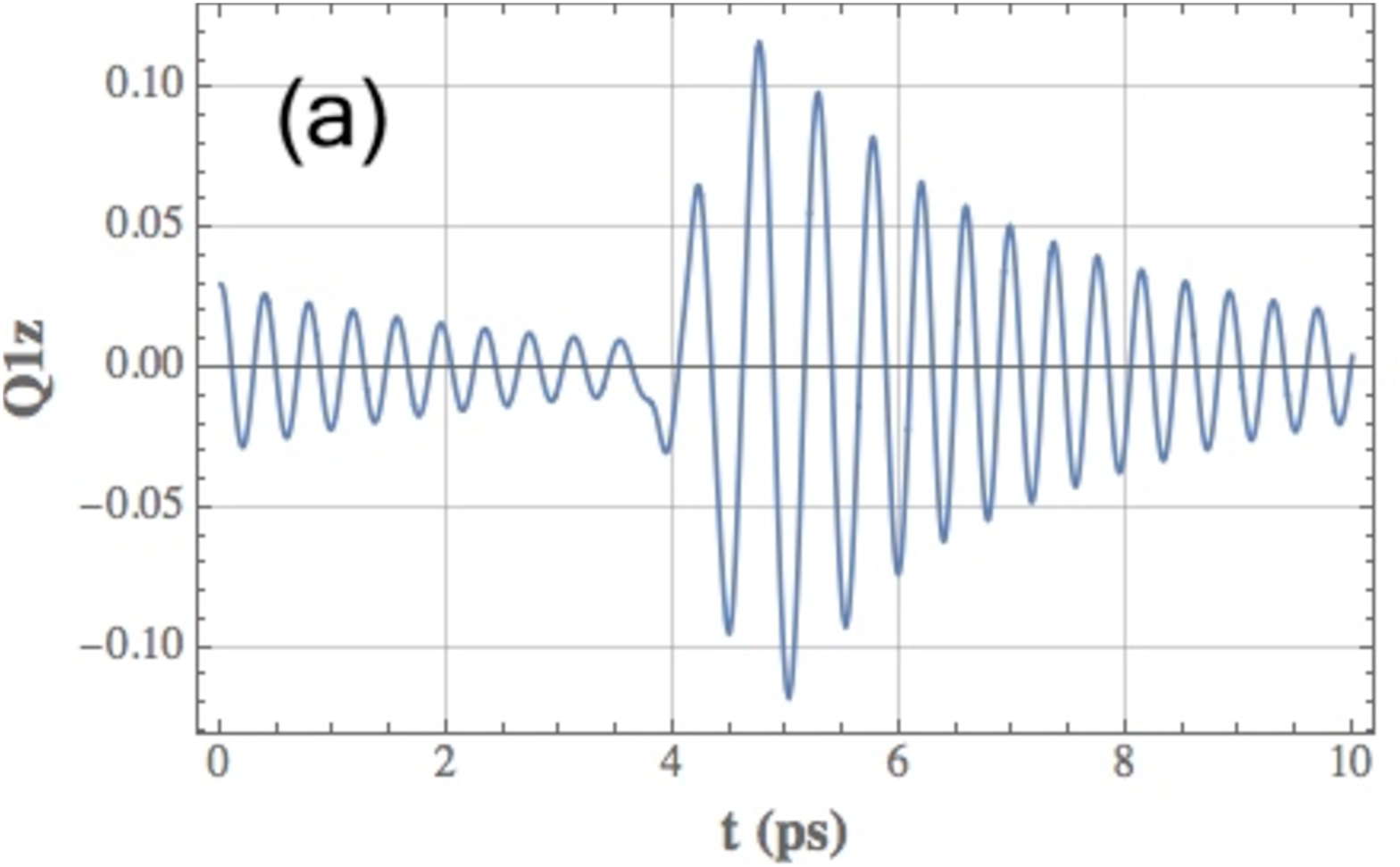}
\includegraphics[width=42mm]{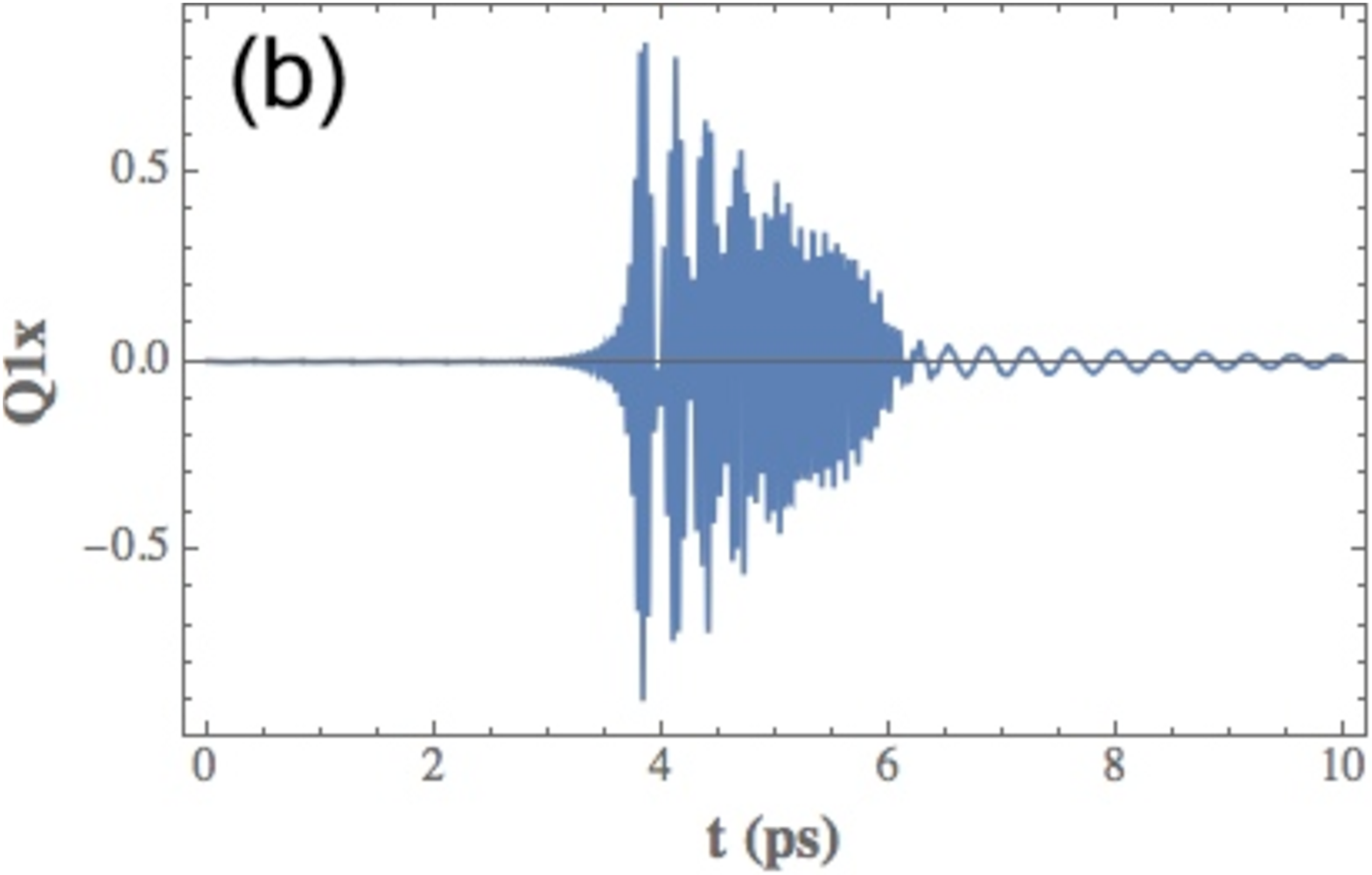}
\includegraphics[width=42mm]{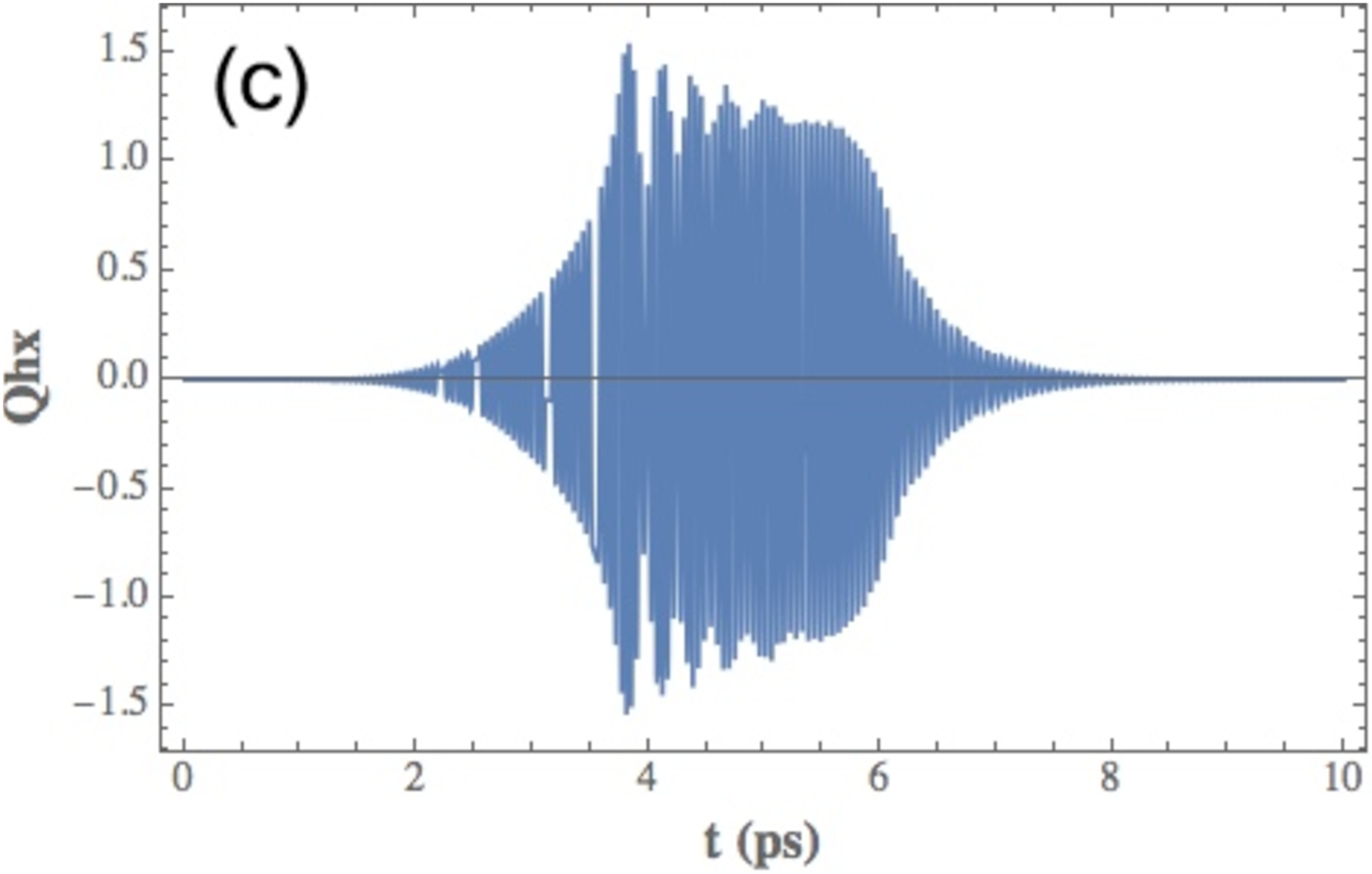}
\includegraphics[width=42mm]{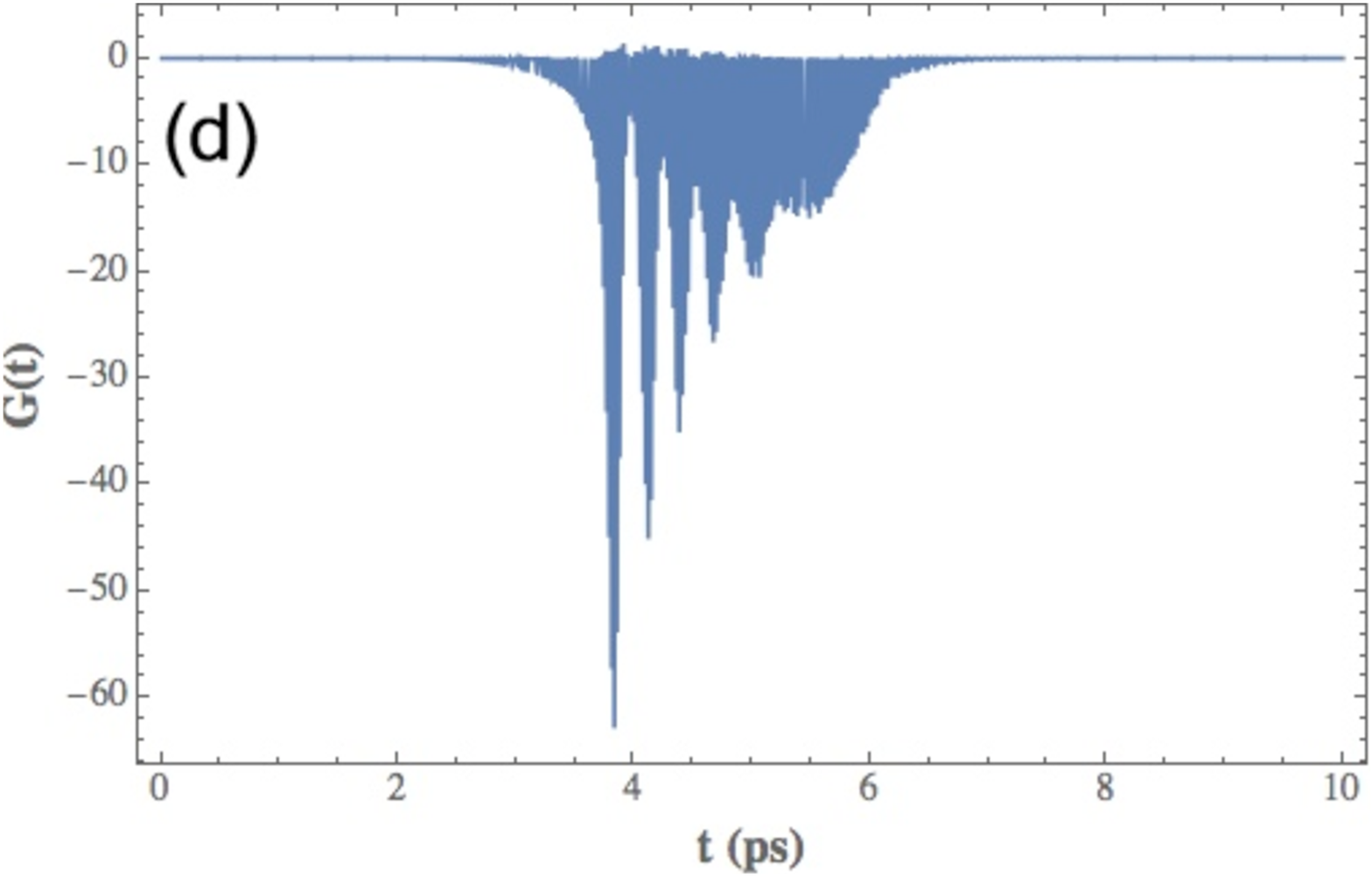}
\caption{ Dynamics of the system under off-resonant driving with frequency $\Omega=1.2 \Omega_h$.  From top to bottom:  $Q_{1z}, Q_{1x},Q_{hx}$ modes, and the instanteneous coefficient $G(t)$ of the $(\frac{\Omega_1^2}{2}+G(t))Q_{1z}^2$ term in the  potential energy. The form of the electric field pulse is the same as in Fig.[\ref{E30}] (but with shifted base frequency), and  $Q_{1z}$ mode is now remarkably excited. When  $G(t)$ exceeds $-\Omega_1^2 \approx -27.06$,  the effective potential for the mode $Q_{1z}$  becomes unstable. Due to violent beatings in $Q_{hx}$ mode only for a short fraction of the pulse length the mode $Q_{1z}$ experiences the inverted parabolic potential.  \label{offresonant}}
\end{figure}
This Hamiltonian is often encountered in problems of celestial mechanics and plasma physics \cite{AKN}.
Under slow change of frequency and/or amplitude of driving, parameters of  (\ref{Hres}) are  changing (increasing frequency corresponds to $\dot \lambda >0$).  Corresponding phase portraits are shown on  Fig. (\ref{duffFig}). A bifurcation happens at  $\lambda_* = \frac{3}{2} \mu^{2/3}$. Below this value,  there is a single equilibrium ($A$), while at higher values of $\lambda$ there are two stable ($A,B$) and one unstable equilibrium ($C$).  Provided certain conditions are met, a phase particle can follow the initial equilibrium point  ($A$) which moves away from origin (correspondingly, in the original system amplitude of oscillation grows, while its frequency remains approximately equal to the instanteneous frequency of the drive,  so this regime is called capture into the resonance ).
 Under influence of a gaussian pulse with fixed frequency,  the point  $A$ is shifted by a certain amount and then returns back to the origin (see Fig. \ref{diagram2}b, upper curve). In contrast,  a pulse with sweeping frequency can shift the equilibrium far away (Fig. \ref{diagram2}b, bottom curve).  In the adiabatic approximation, dynamics can be described in a great detail \cite{AKN,ANeishtadt}. E.g., as the parameters of the system are changing,  phase space area within the trajectory remains approximate adiabatic invariant, and from behaviour of the areas inside separatrix loops it can be predicted when the phase point will be thrown away from the resonance. In our realistic system non-adiabaticity and dissipation become very important, nevertheless qualitative understanding of dynamics allows to construct simple and effective protocol for phonon excitation. With dissipation, the equation of motion becomes 
 \be
\ddot{Q} + \gamma \dot{Q} + \Omega_0^2 Q + 4c_4 Q^3 = F_0 \sin{ \Phi(t)}  \label{ duff},
\ee
which is the same as in a damped and driven Duffing oscillator. At fixed frequency, searching for a periodic solution $Q = A \sin (\Omega t + \phi_0)$,  one gets a cubic equation for the amplitude of a steady solution:
\be
A^2  \Bigl( \gamma^2 \Omega^2 + ( \Omega_0^2 - \Omega^2+ 3 c_4 A^2 ) ^2 \Bigr)   = F_0 ^2   \label{ampl}
\ee
Solutions of Eq.(\ref{ampl}) are shown on Fig. (\ref{duffFig}a)  for different values of the driving amplitude $F_0$. They are dissipative counterparts of 
 fixed points in phase portraits of  (\ref{Hres}). One can see that driving frequency higher than the resonant one allows to achieve higher steady-state amplitudes.
\begin{figure}
\includegraphics[width=42mm]{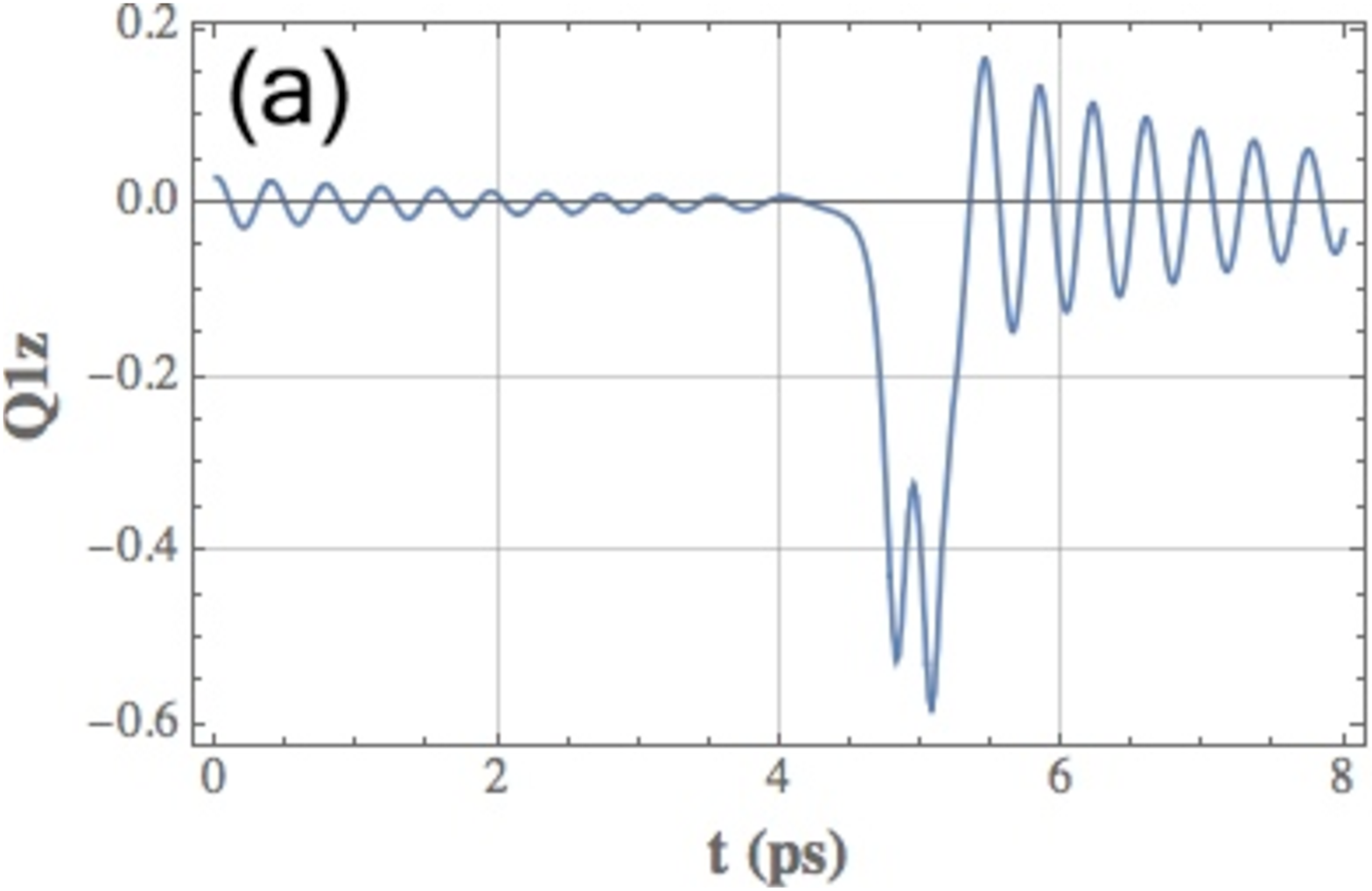}
\includegraphics[width=42mm]{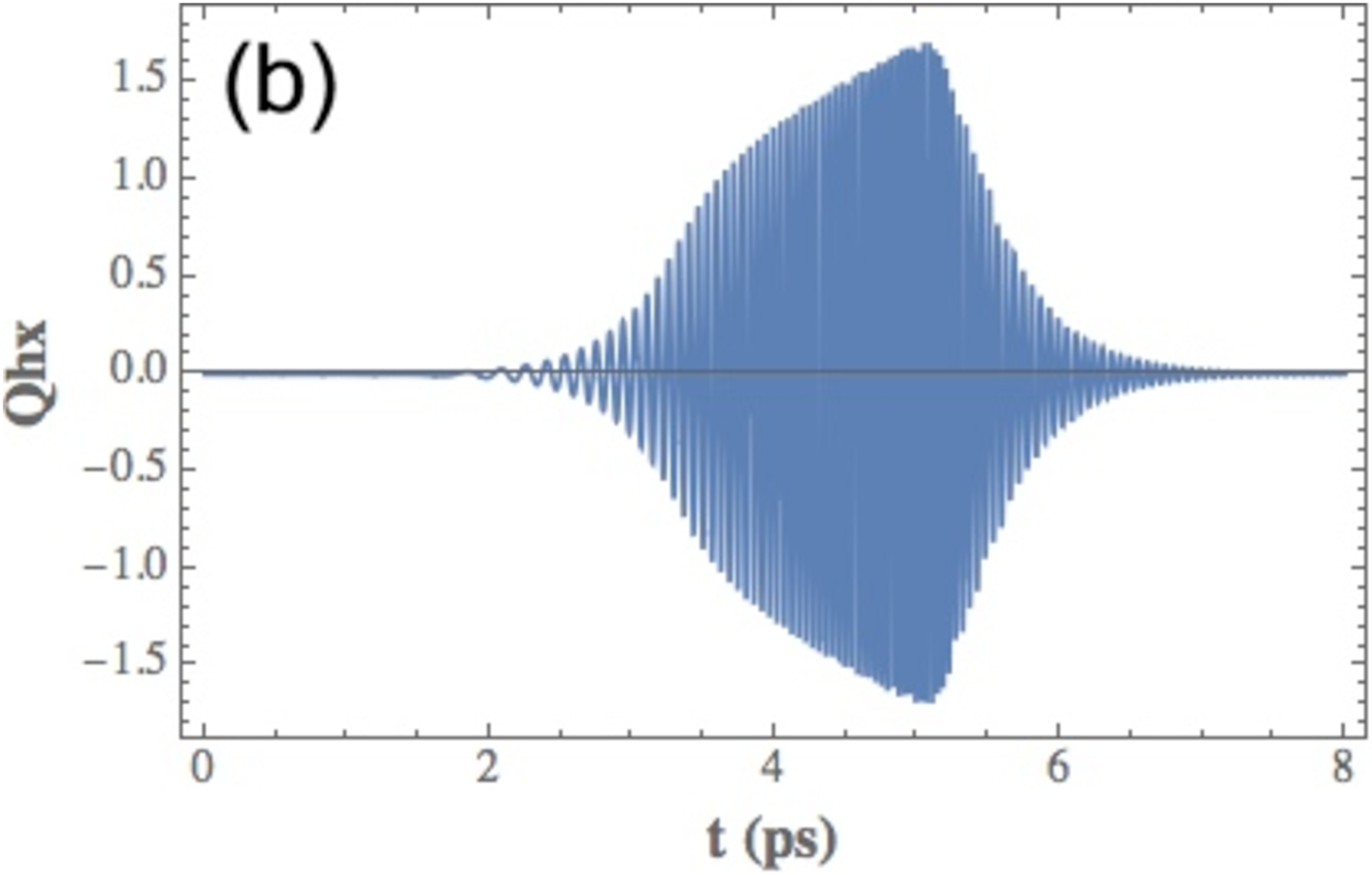}
\includegraphics[width=42mm]{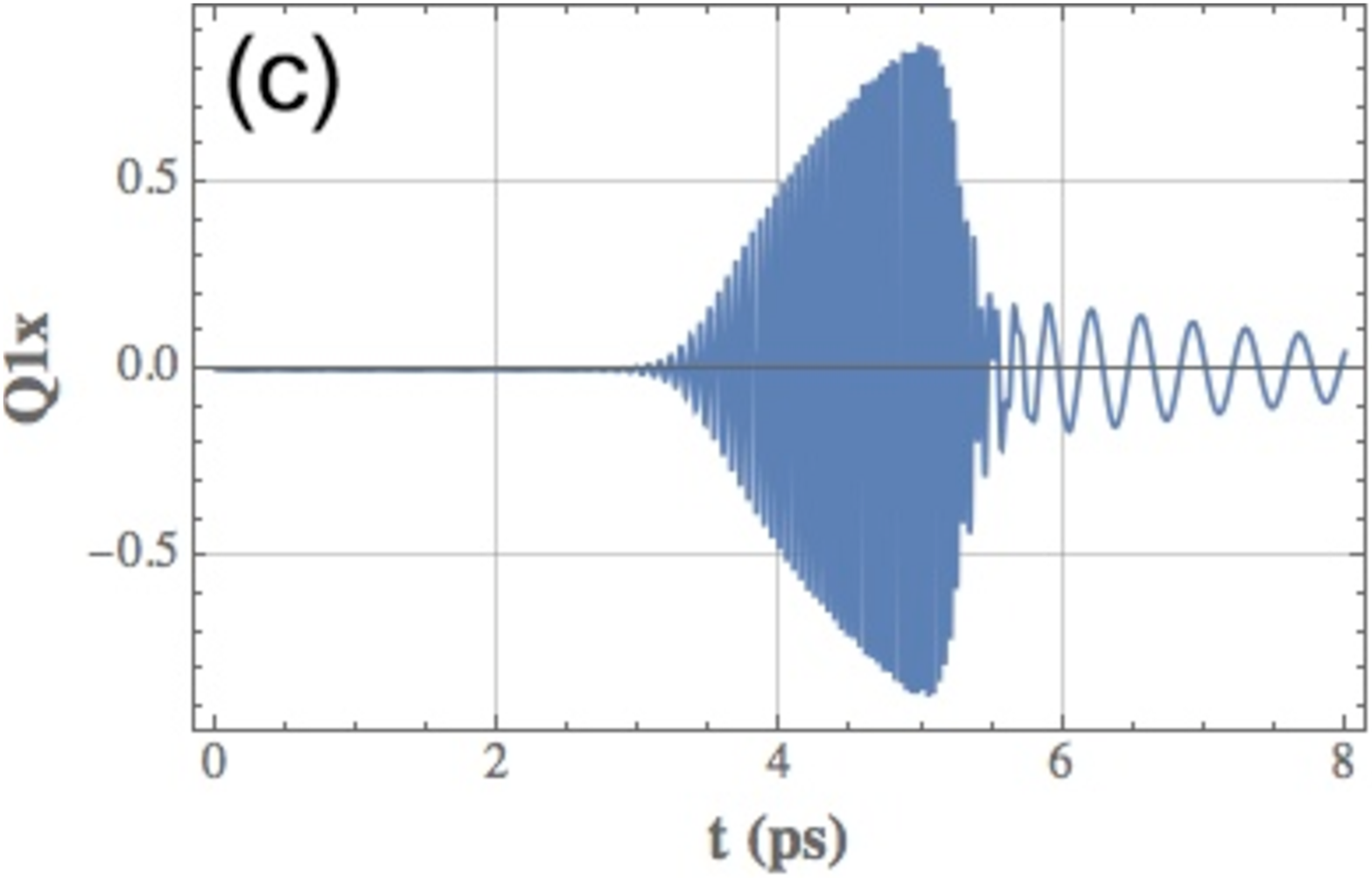}
\includegraphics[width=42mm]{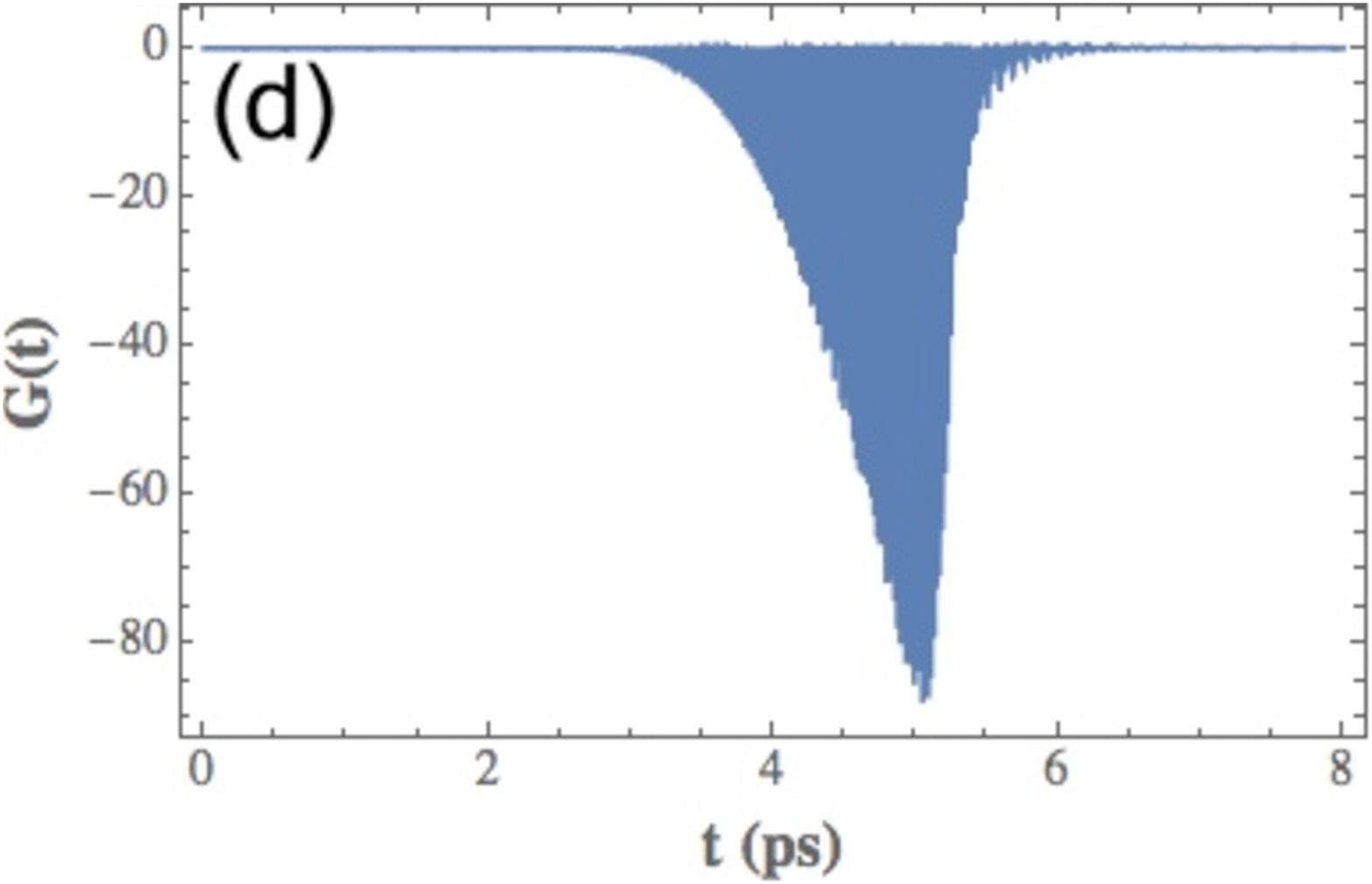}
\includegraphics[width=42mm]{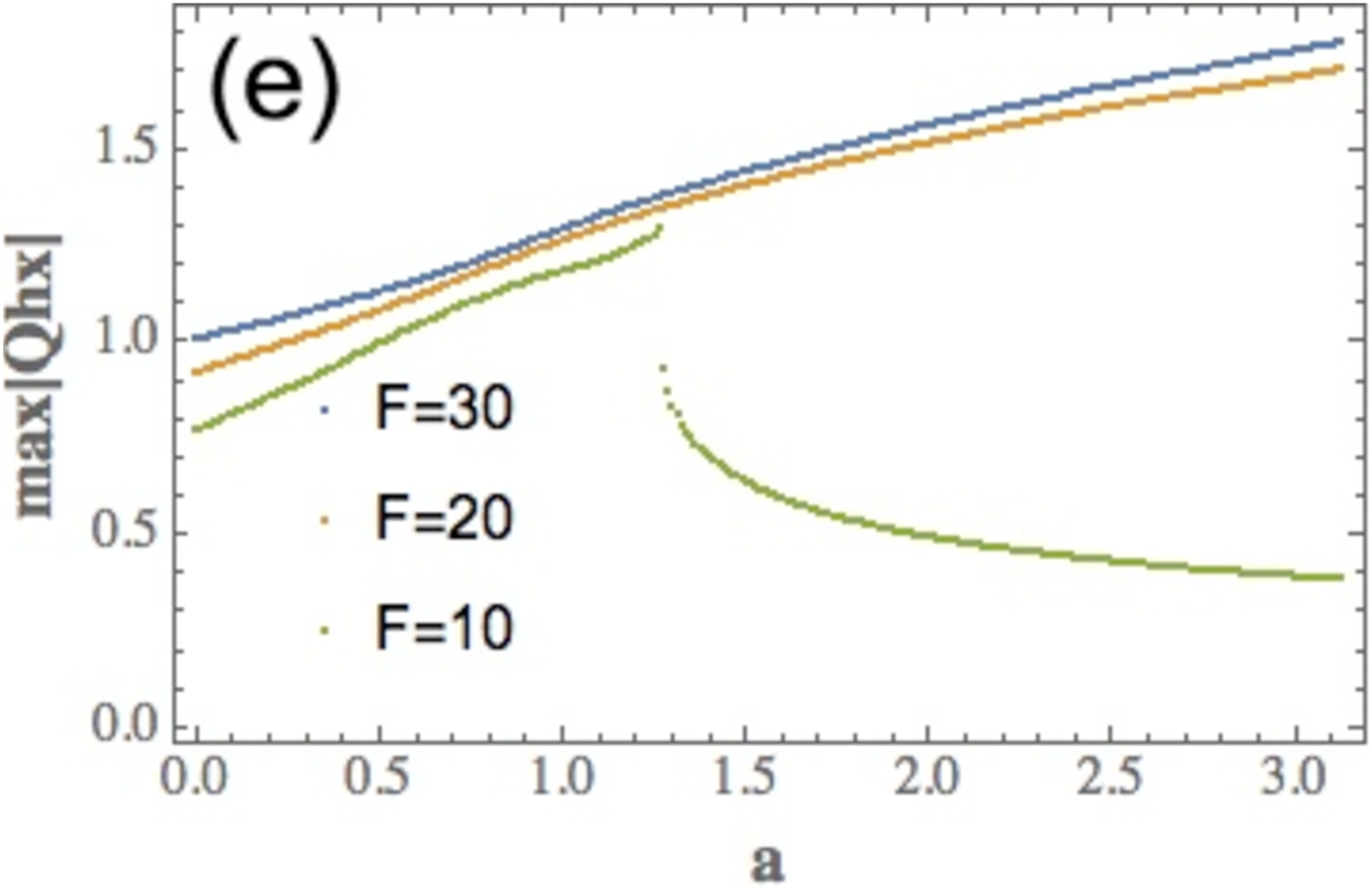}
\includegraphics[width=42mm]{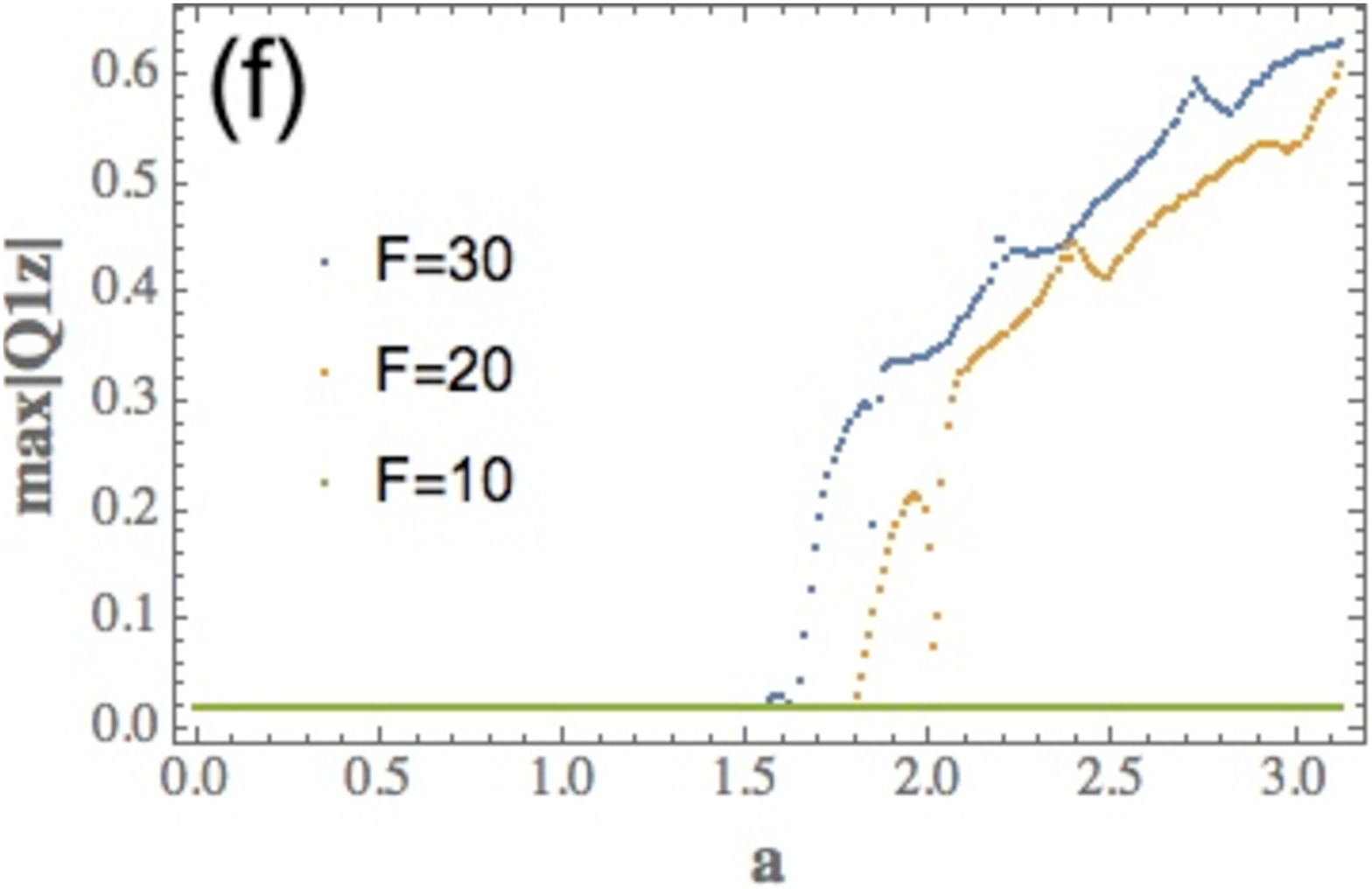}
\caption{ Dynamics of the KTa$O_3$ model system under  driving with sweeping frequency. Instanteneous frequency in the center of the pulse is $\Omega=1.2 \Omega_h$. (a-d)    $Q_{1z}, Q_{1x},Q_{hx}$ modes, and the instanteneous coefficient $G(t)$ of the $(\frac{\Omega_1^2}{2}+G(t))Q_{1z}^2$ term in the  potential energy.   (e,f)  Maximal amplitude of (e) $Q_{hx}$ and (f) $Q_{1z}$ modes under  driving with sweeping frequency at different sweeping rates $a$.  Curves from top to bottom correspond to amplitudes of driving $F_0=30,20,10$ MV cm$^{-1}$. Damping rates are 5\% of the linear frequencies.  Instanteneous frequency of the drive is linearly increased, reaching the linear frequency of the phonon mode $\Omega_{h}$ at $t_*=0.75 \sigma$ before the maximum of the pulse. $\sigma = 2$ ps.  $a=0$ corresponds to the drive with the constant frequency $\Omega=\Omega_{h}$.  \label{sweep}}
\end{figure}
\begin{figure}
\includegraphics[width=42mm]{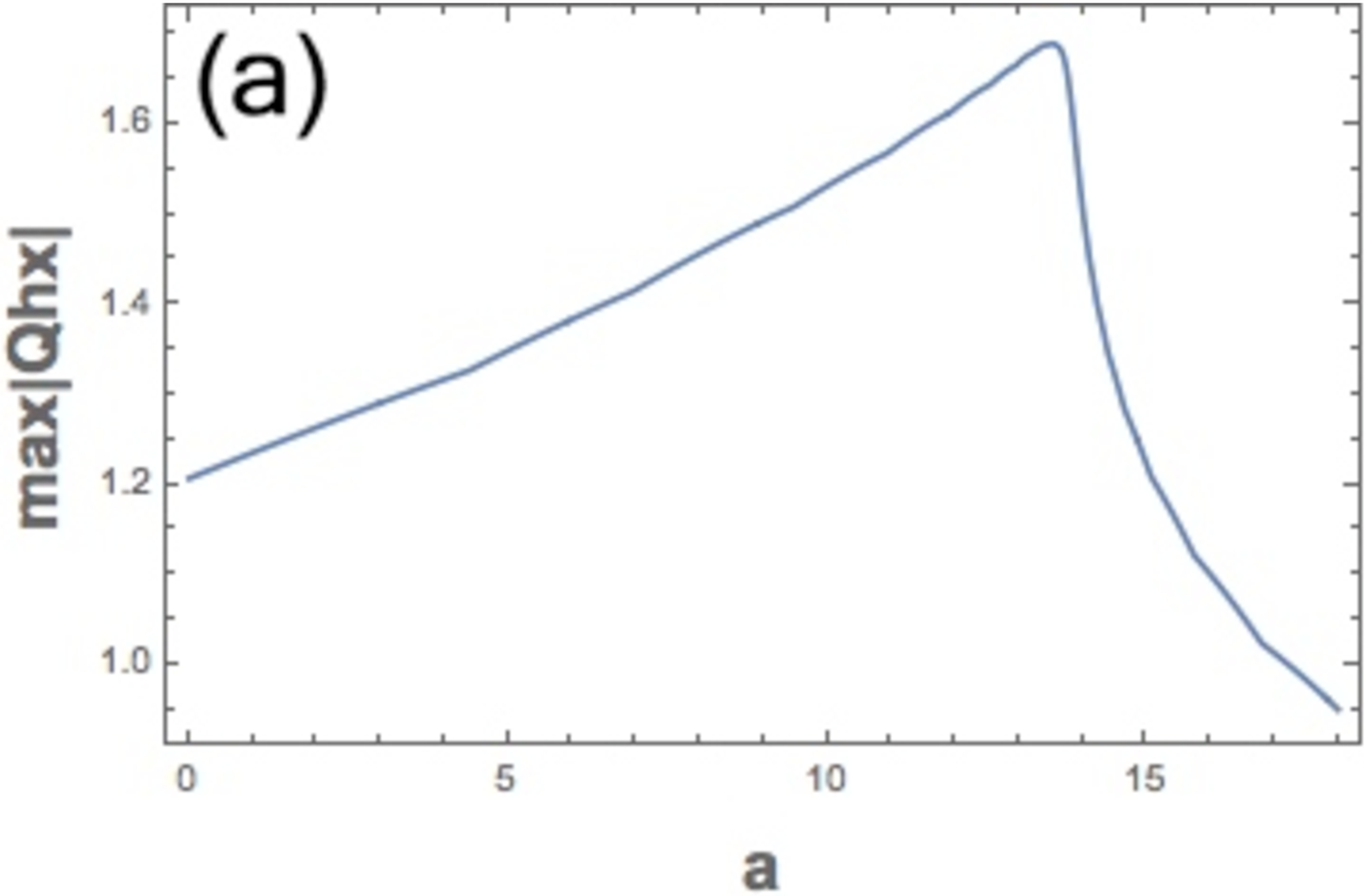}
\includegraphics[width=42mm]{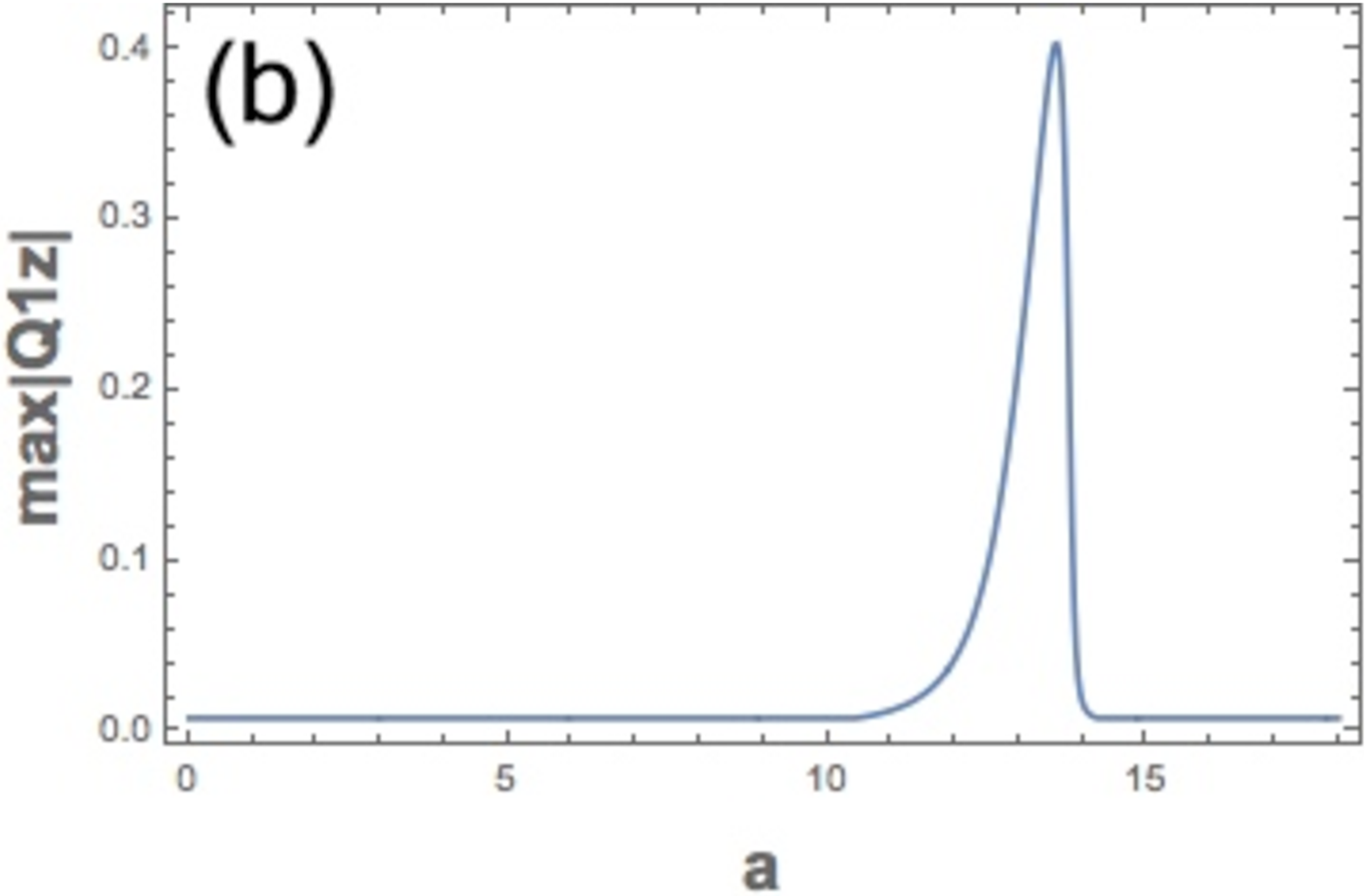}
\includegraphics[width=42mm]{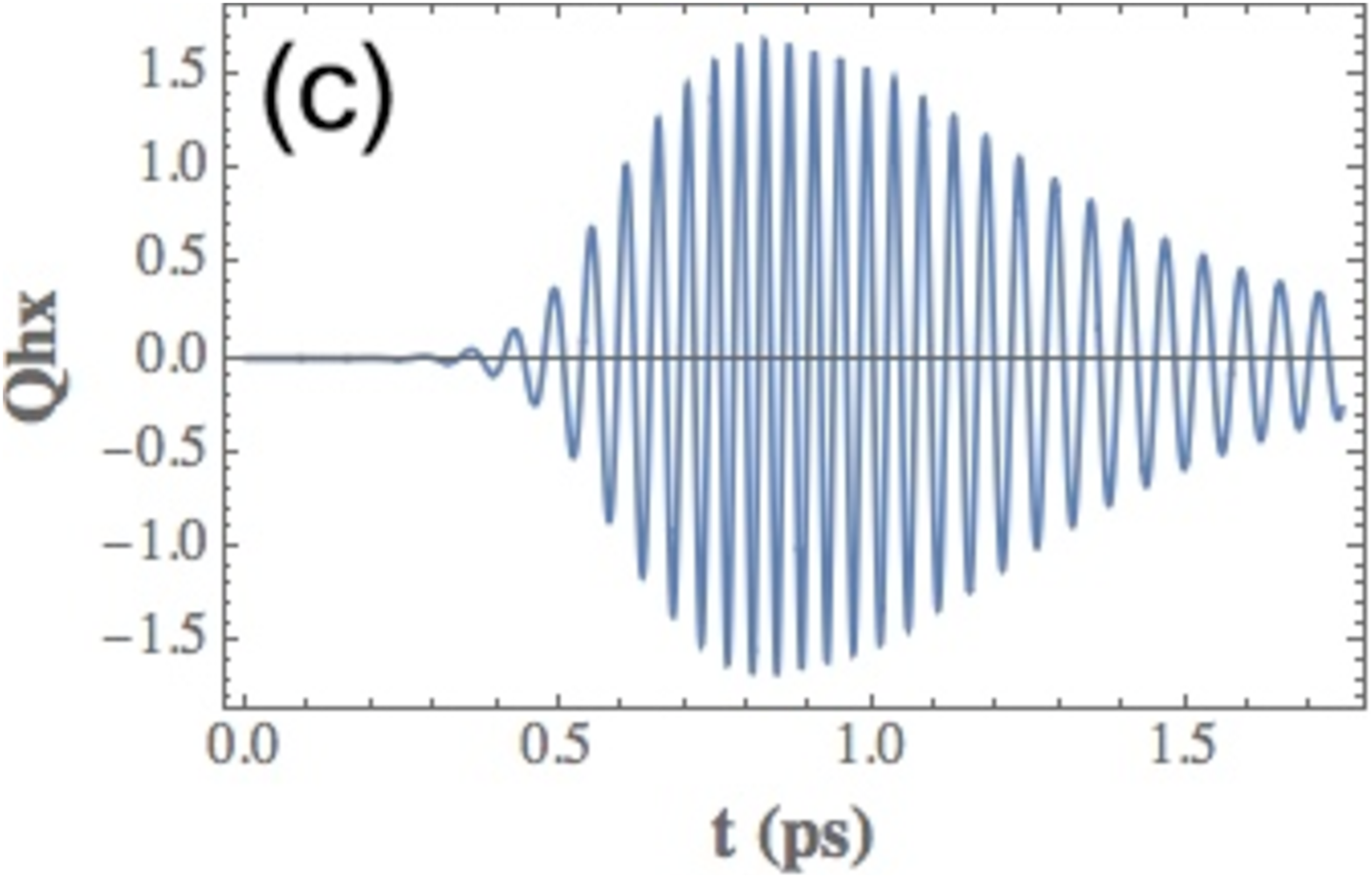}
\includegraphics[width=42mm]{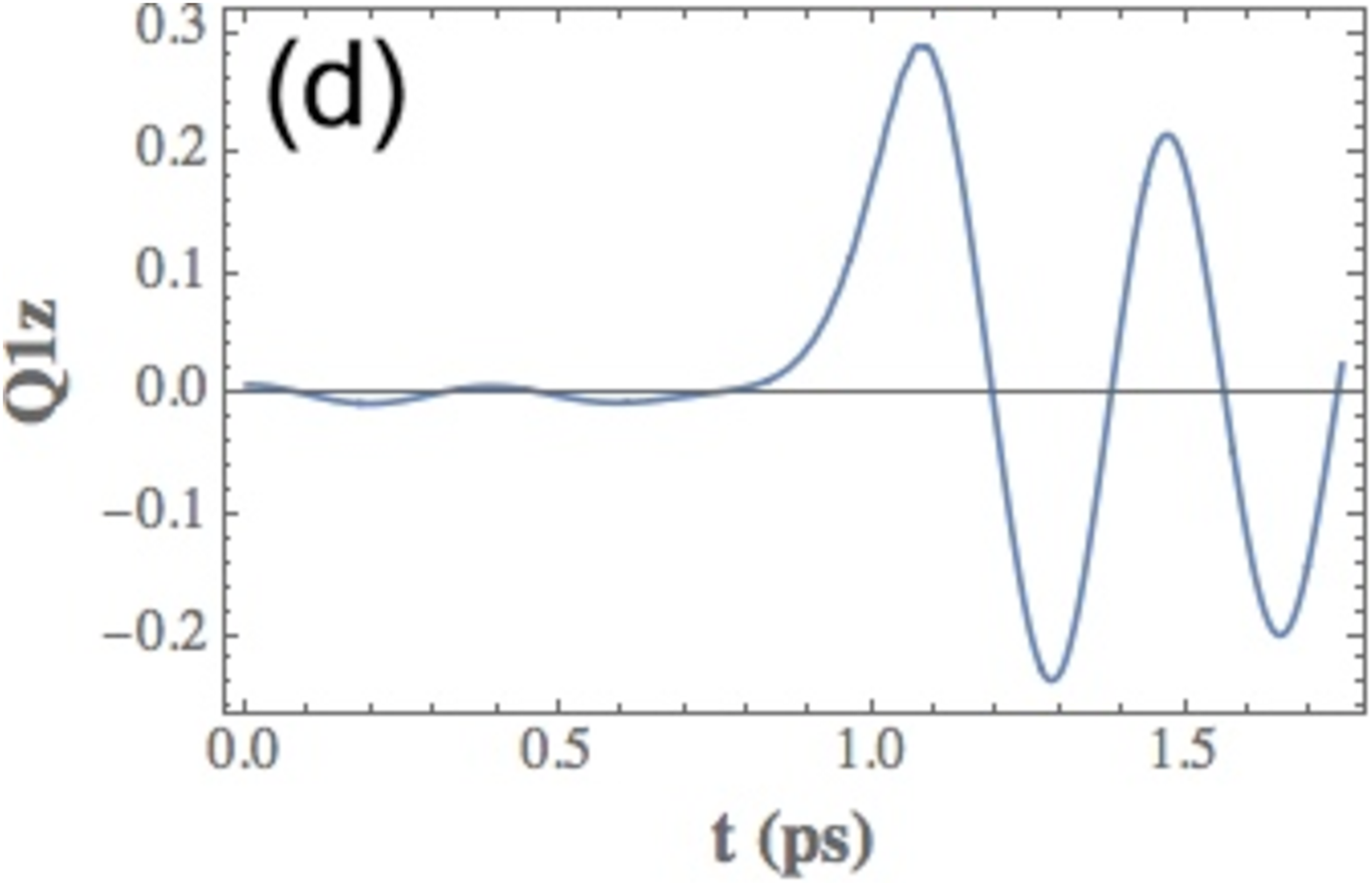}
\caption{Excitation by a 350-femtosecond pulse. (a,b) Maximal amplitude of modes $Q_{hx}, Q_{1z}$ under  driving with sweeping frequency at different sweeping rates $a$. (c) Mode $Q_{1z}$ at a=13.2 (d)  Mode $Q_{hx}$ at a=13.2.  The pulse width is $\sigma = 0.35$ ps.   \label{diagram1}}
\end{figure}
To demonstrate induced ferroelectricity, consider a specific example of  KTaO$_3$:  a perovskite oxide with cubic structure,  posessing a paraelectric phase. Such materials have four triply degenerate optical phonon modes at the zone center. Three of these modes are infrared active (have the irreducible representation $T_{1u}$ \cite{SubediFerro}). The remaining one is optically inactive ( \cite{SubediFerro}).  Ferroelectricity is related to  dynamical instability of an infrared-active transverse optic phonon mode:  most ferroelectric
materials show a characteristic softening of an infrared transverse optic mode as the transition temperature is approached. 
In \cite{SubediFerro} it was investigated if a similar softening and instability of the lowest frequency $T_{1u}$ mode
can be achieved by an intense laser-induced excitation of the highest frequency $T_{1u}$ mode.
In the case of cubic structure it was not achieved due to certain dynamical reasons (being discussed below),  however with addition of internal stress which modifies the crystal lattice such mechanism worked. Let us  consider in a more detail the cubic structure case (without stress).  The calculated in \cite{SubediFerro}   phonon frequencies are   $\Omega_1=85$ cm$^{-1}$ and $\Omega_h=533$ cm$^{-1}$ for the lowest and highest frequency $T_{1u}$ modes, respectively ($\Omega_0$ from previous discussion plays the role of  $\Omega_h$ now).
Following  \cite{SubediFerro} we   simplify  analysis  by considering a  case where the $x$ component of the highest frequency  mode $Q_{hx}$ is pumped by an intense light source 
and influences the dynamics of the lowest frequency $T_{1u}$ modes along the longitudinal $Q_{1x}$ and transverse $Q_{1z}$ coordinates.
Dynamics along the second transverse coordinate $Q_{1y}$ is neglected. The energy surface $V(Q_{hx},Q_{1z},Q_{1x})$ has a complicated form with many kinds of 
nonlinear couplings and anharmonicities:
$ V(Q_{hx},Q_{1z},Q_{1x}) + \frac{\Omega_1^2}{2} ( Q_{1z}^2 +  Q_{1z}^2 ) + \frac{\Omega_h^2}{2} Q_{hx}^2 + V^{nl}(Q_{hx},Q_{1z},Q_{1x}),  $
 where $ V^{nl}$ is the nonlinear part of the energy, obtained in state-of-the-art calculations of \cite{SubediFerro}, and given in \cite{Suppl} for convenience.
Equations of motions are:
\bea
\ddot{Q}_{hx}  + \gamma_h \dot{Q}_{hx}  + \Omega_{hx}^2 Q_{hx} &=& -\frac{\partial{ V^{nl}(Q_{hx},Q_{1z},Q_{1x})}}{\partial{Q_{hx}}} +F(t), \nonumber\\
\ddot{Q}_{1j}  + \gamma_1 \dot{Q}_{1j}+ \Omega_{1}^2 Q_{1j} &=& -\frac{\partial{ V^{nl}(Q_{hx},Q_{1z},Q_{1x})}}{\partial{Q_{1j}}}, \label{system}
\eea
where $j =x,z$; $\gamma_h$ and $\gamma_1$ are damping coefficients (typically few percents of the corresponding harmonic frequencies),  external force $F(t) = Z_{hx}^* E_0 \sin(\Omega t) e^{-t^2/2\sigma^2} $, $Z_{hx}^*$ is the effective charge and $\Omega$ is the driving frequency.  
\begin{figure}
\includegraphics[width=42mm]{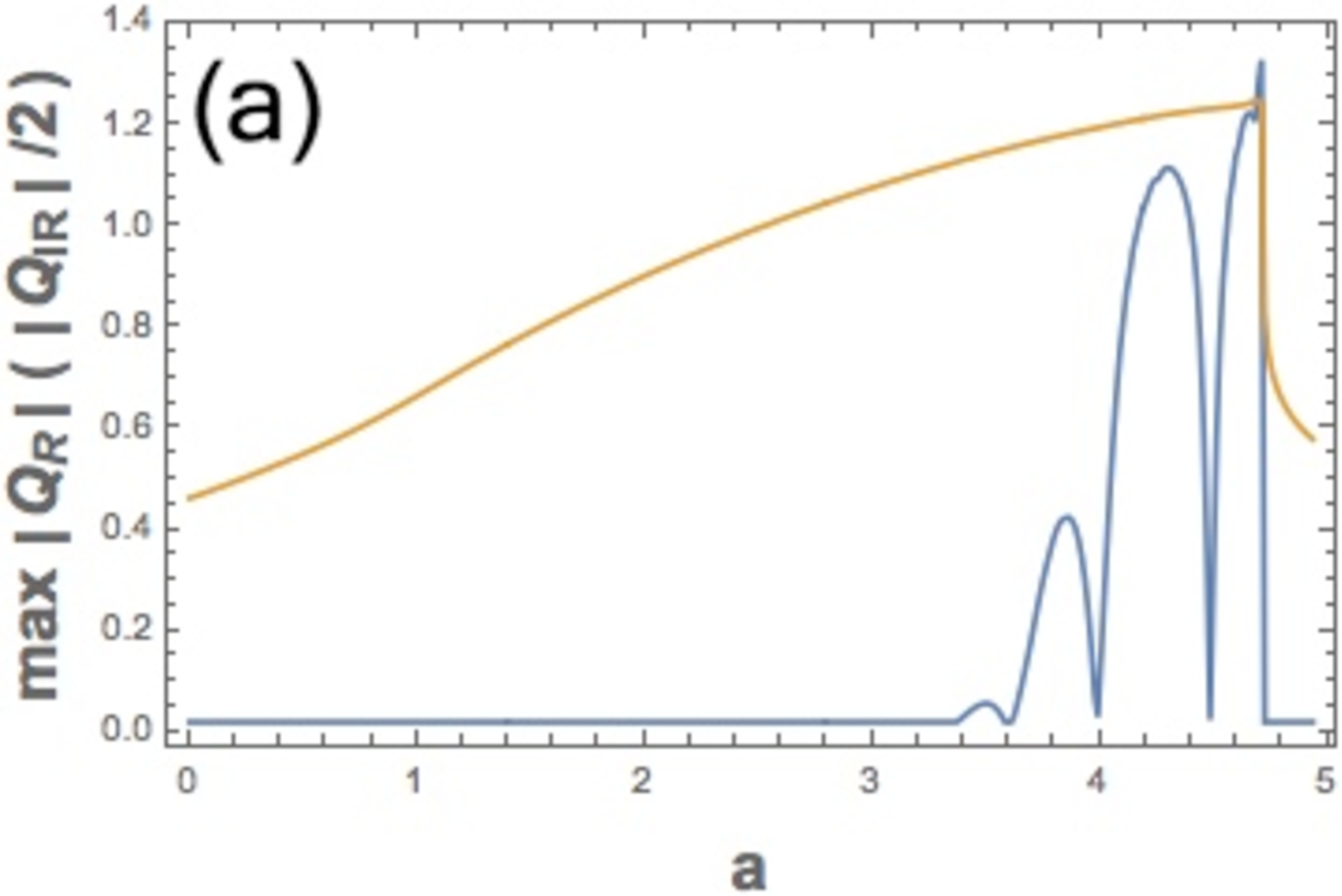}
\includegraphics[width=42mm]{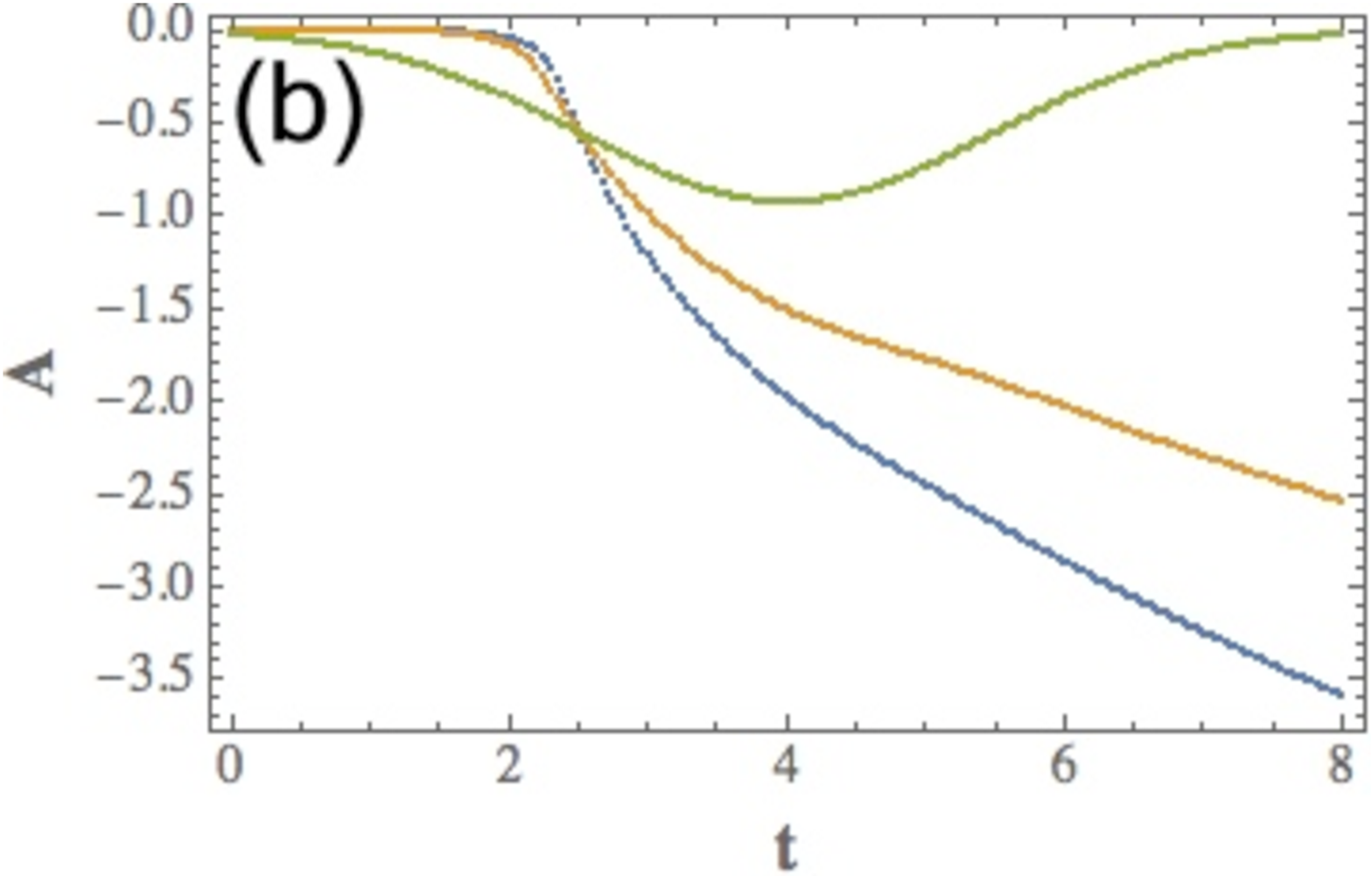}
\caption{ (a) Maximal amplitude of modes $Q_{IR}, Q_{R}$ in the driven LCO system under driving with sweeping frequency at different sweeping rates $a$ (for clarity  $Q_{IR}/2$ is shown (top curve)  )  (b) corresponding properties of the effective Hamiltonian (Location of the stable equilibrium of effective potential as a function of time for 3 different values of sweeping rates. Only the lowest curve lead to excitation of the Raman mode. ).  $\sigma = 2$ ps.     \label{diagram2}}
\end{figure}
Qualitative understanding of possible dynamics can be captured by drawing a projection of the potential energy $V(Q_{1z}, Q_{1x}, Q_{hx})$  by the plane $Q_{1x}=0$  (Fig. 3 of \cite{SubediFerro}).
Resulting curves $V(Q_{1z})$ at fixed values of $Q_{hx}$ have single-well form (at small absolute values of $Q_{hx}$), or a double well form (at larger $Q_{hx}$).  In the latter case,  induced ferroelectricity is possible, as finite value of $Q_{1z}$ at equilibrium corresponds to the ferroelectric phase.  However, to reach such a state dynamically is a nontrivial issue. Excitation of  $Q_{hx}$ mode should be done with a pulse of limited power and duration. 
When the driving frequency $\Omega$ is chosen close to the phonon mode frequency $\Omega_h$ \cite{SubediFerro},  the transverse mode $Q_{1z}$ remains almost unaffected (Fig. (\ref{E30}) due to insufficient amplitude of  the $Q_{hx}$ mode. In \cite{SubediFerro}, the amplitude of driving was varied in a large  range, up to pump amplitudes of $100 MV cm^{-1}$, with no signs of dynamical instability in $Q_{1z}$. 
Increase of the pump amplitude  makes the dynamics of $Q_{hx}, Q_{1x}$ chaotic, but  does not result in noticeable response of  the $Q_{1z}$ mode:  at strong driving the 'axuilary' longitudinal mode $Q_{1x}$ becomes excited (\cite{Suppl}), and  chaotic dynamics prevents efficient excitation of $Q_{hx}$. There is a range of driving frequencies (away from the exact resonance with $\Omega_h$), where a pulse of the same amplitude can effectively excite all three modes, inducing transient dynamical instability in $Q_{1z}$. Indeed,  shifting driving frequency about 15-20\%  from the exact resonance with  $\Omega_h$  leads to considerable excitation of the modes  $Q_{1z}$, $Q_{hx}$  (Fig. (\ref{offresonant}) ).  The mode $Q_{hx}$  experiences beatings which create transient double-well potential  for the $Q_{1z}$ mode (Fig. (\ref{offresonant})d).  From the full potential energy of the system, we can single out the term quadratic in $Q_{1z}$ : 
 $\left( \frac{\Omega_1^2}{2}+G(t) \right) Q_{1z}^2$, where $G(t)=  m_2 Q_{hx}^4 + d Q_{hx} Q_{1x} + l Q_{hx}^2 + 
 p  Q_{1x}^2  + e_2  Q_{hx}^2 Q_{1x}^2 + e_3  Q_{hx}^3 Q_{hx}$ ( \cite{Suppl} ). Due to strong coupling between $ Q_{hx}$ and $ Q_{1x}$ modes, they oscillate synchronously although their linear frequencies are very different. When the average value of the coefficient $ \langle G(t) \rangle$ exceeds $-\Omega_1^2/2 $,  the effective potential for the mode $Q_{1z}$  becomes unstable. Due to violent beatings in the $Q_{hx}$ mode, it does not happen smoothly, and only for a short fraction of the pulse excitation time the mode $Q_{1z}$ experiences the inverted parabolic potential (at the first maxima of the beatings, see Fig. (\ref{offresonant})d). 
There is a way to create the needed effective potential in a more robust way. Consider  driving with sweeping frequency. A chirped pulse has the form $F= F_0(t) \sin \Phi(t)$, where   $ \Phi(t) = \Omega_0 t +  \frac{\alpha t^2}{2},$ $F_0(t) = \exp \left( -t^2/2(\sigma / 2 \mbox{ln} 2)^2 \right).$   Time-dependence of the instantenious frequency and the amplitude translates into the dependence of parameters $\mu, \lambda$ of the Hamiltonian $(\ref{Hres})$ on time.  Corresponding phase portraits are slowly deformed,  and, if our phase point is not thrown away from the region where the initial equilibrium is located (see \cite{ANeishtadt} for details), it oscillates around the equilibrium point moving away from the origin.
Such regime, illustrated in Fig.(\ref{sweep}), is not only effective for excitation of the $Q_{hx}$ mode, but also provides smooth generation of the effective potential  for the $Q_{1z}$ mode (Fig.(\ref{sweep})d).
The important feature of the dynamics is that the axillary longitudinal mode  $Q_{1x}$  also gets excited considerably. Unlike the case of very strong resonant driving (\cite{Suppl}), where  chaotic dynamics happens after excitation of the $Q_{1x}$ mode, here the longitudinal modes oscillate synchronously (in 1:1 resonance), and the resulting dynamics is regular.
Regular dynamics happens because the pulse with sweeping frequency excites the system in such a way that a phase point remains not far from the instanteneous equilibrium point.   

We note that damping play important role in dynamics. Damping parameters typically are 5-10 \% of corresponding linear frequencies. In case we assume higher damping for low-frequency phonons,  the axillary longitudinal mode  $Q_{1x}$  is not excited considerably, and dynamics can be understood from the driven Duffing oscillator model for the $Q_{hx} $ mode alone (with small corrections from $Q_{1x}$).  While in the Hamiltonian model there are three equilibria above the critical frequency detuning, and a phase point  captured into resonance can reach large amplitudes of $Q$ moving near one of them, damping leads to termination of this process at the tip of the resonance 'tongue', where stable and unstable equilibria collide and annihilate.  At weak driving amplitudes,  the tips lie at a 'backbone' defined as $A_{tip} = \frac{F_0}{\gamma \Omega_0}$,  $\omega_{tip} =  1 + \frac{3 c_4 F_0^2} {\gamma^2 \Omega_0^4} $.  For a rough estimate of the optimal sweep, assume that a pulse starts with the linear resonance frequency $\Omega_0$,  and reaches the tip of the resonance 'tongue' at its maximum. Then, the estimate for the sweeping rate is $
\alpha =  \frac{3c_4 F_0^2}{\gamma^2 \Omega_0^3}.$ We make numerical experiments with various sweeping rates and amplitudes of driving (see Fig.\ref{sweep}e,f), 
and find a remarkable improvement in efficiency of excitation compared to pulses with constant frequency.  Most exciting, the protocol with sweeping frequency works also for much shorter, sub-picosecond pulses. We show in  Fig.\ref{diagram1} an example with $\sigma$=350fs,  which corresponds to laser parameters used in A.Cavalleri  group.

The same approach applies not only to ferroelectrics, but to many other systems, e.g. laser driven LCO  \cite{Phononics}.
There,  a relevant reduced model consists of infrared-active $Q_{IR}$ mode (described by a driven Duffing oscillator) coupled to a Raman mode $Q_{R}$ by a quadratic-quadratic term.  $Q_R$ ($B_{1g}(18)$) mode describes in-plane rotations of CuO$_6$ octahedra, whereas $Q_{IR}$ mode  describes in-plane stretching of Cu-O bond. 
The energy surface has the form $
V =  \Omega_0^2  \frac{Q_{IR}^2}{2}  +  \Omega_1^2  \frac{Q_{R}^2}{2}  + c_4  Q_{IR}^4 + b_4 Q_{R}^4  - \frac{g}{2}  Q_{IR}^2 Q_{R}^2. $
Values of the coefficients were derived in elaborate calculations of  \cite{Phononics}.
There is a single-potential well around the equilibrium value for $Q_R$ mode at small amplitudes of $Q_{IR}$, which becomes double-well potential at larger amplitudes of $Q_{IR}$ (instanteneous quadratic potential felt by the slow mode is $\frac{Q_{R}^2}{2}  \left(  \Omega_{1}^2 -  g  Q_{IR}^2 \right)$, which becomes inverted parabolic potential for sufficiently high amplitudes of the driven $IR$ mode.  The critical value of driving force $F_c$ depends on detuning $ \delta \Omega \equiv \Omega-\Omega_{0} $ and can be made smaller than its value on the resonance (being used in \cite{Phononics} ). 
Indeed, to the first approximation,  the averaged potential for the slow mode is $\frac{Q_{R}^2}{2}  \left(  \Omega_{1}^2 -  g  Q_{IR,max}^2/2 \right)$ and becomes unstable at critical value of the fast mode amplitude
$Q_{IR,max}=   \Omega_{1} \sqrt{2/g} $.  This can be achieved at sufficiently smaller driving force amplitudes provided sweeping frequency pulse is used. We show corresponding results of numerical calculations in Fig.(\ref{diagram2}). Fig.(\ref{diagram2})c shows also instanteneous locations of  stable equilibrium of the effective potential as a function of time for  different values of sweeping rates. 

Excitation of the  in-plane rotations associated with the $Q_{R}$ mode can be used to modulate superexchange  coupling in this
cuprate \cite{Phononics}.  Recently there has been a lot of interest in effective models arising from periodic driving \cite{APIK,Polkovnikov,Eckardt, Mentink,Abdullaev,Dalibard},  and the suggested method can be useful for this area of research as well. The proposed method can be useful also for recent proposals and experiments on driven orthorombic perovskites (like ErFeO$_3$, see \cite{Spaldin,Kimel} ), where three-linear phonon coupling is realized:   two high-frequency infrared-active modes are coupled to the third, Raman mode.

To conclude,  we demonstrate that  drastic improvement in efficiency of excitation of nonlinear phonons can be achieved using chirped pulses.  In terms of nonlinear dynamics of reduced
classical models, capture into the resonance happens and the driven mode is transferred to a higher amplitude state efficiently, which triggers instabilities in the coupled low-frequency modes, and corresponding phase transitions. The method is especially remarkable in cases where a system cannot be excited by bare increase of the power of drive, like in KTaO$_3$.
The approach can be useful in many recent proposals on laser-induced phase transitions, including induced ferroelectricity in perovskites, induced structural transitions in LCO cuprates, and excitation of orthorombic perovskites.

We are grateful to  A.I.Neishtadt and A.Cavalleri for insightful discussions.
The work was supported by  NWO via Spinoza prize and by European Research Council (ERC) Advanced Grant No. 338957 FEMTO/NANO.

\widetext
\newpage
\begin{center}
\vskip 5cm
\textbf{\large Supplemental Materials: Efficient excitation of nonlinear phonons via chirped mid-infrared pulses: induced structural phase transitions  }
\end{center}
%%%%%%%%%% Merge with supplemental materials %%%%%%%%%%
%%%%%%%%%% Prefix a "S" to all equations, figures, tables and reset the counter %%%%%%%%%%
\setcounter{equation}{0}
\setcounter{figure}{0}
\setcounter{table}{0}
\setcounter{page}{1}
\makeatletter
\renewcommand{\theequation}{S\arabic{equation}}
\renewcommand{\thefigure}{S\arabic{figure}}
\renewcommand{\bibnumfmt}[1]{[S#1]}
\renewcommand{\citenumfont}[1]{S#1}
%%%%%%%%%% Prefix a "S" to all equations, figures, tables and reset the counter %%%%%%%%%%

\subsection{Nonlinear part of the potential energy surface for \mbox{KTaO}$_3$}

The nonlinear part of the potential energy surface is:
\bea
V^{nl}(Q_{hx},Q_{1z},Q_{1x}) &=&  \\  =\sum \limits_{k=2}^6 \Bigl( a_{2k} (Q_{1z}^{2k} &+& Q_{1x}^{2k})  + c_{2k}Q_{hx}^{2k}  \Bigr)   \nonumber\\ +l Q_{1z}^2 Q_{hx}^2 &+&m_1  Q_{1z}^4 Q_{hx}^2  + m_2   Q_{1z}^2 Q_{hx}^4 \nonumber\\ +
n_1  Q_{1z}^4 Q_{hx}^4 &+& n_2  Q_{1z}^6 Q_{hx}^2 + n_3  Q_{1z}^2 Q_{hx}^6  \nonumber\\ + t_1 Q_{1x}^3 Q_{hx} &+&
t_2  Q_{1x}^2 Q_{hx}^2 + t_3  Q_{1x} Q_{hx}^3  \nonumber\\ + p Q_{1z}^2 Q_{1x}^2 &+& q_1 Q_{1z}^4 Q_{1x}^2 + q_2 Q_{1z}^2 Q_{1x}^4  \nonumber\\ +  r_1 Q_{1z}^4 Q_{1x}^4 &+&
r_2 Q_{1z}^6 Q_{1x}^2 +r_3 Q_{1z}^2 Q_{1x}^6  \nonumber\\ + d  Q_{1z}^2 Q_{1x} Q_{hx}  &+&   g  Q_{1z}^4 Q_{1x}  Q_{hx}  \nonumber\\ +
 \sum \limits_{k=1}^3 e_k Q_{1z}^2 Q_{1x}^{4-k} Q_{hx}^k  &+&  \sum \limits_{k=1}^5 u_k Q_{1x}^{6-k} Q_{hx}^k \nonumber\\ +
 \sum \limits_{k=1}^5 f_k Q_{1z}^2 Q_{1x}^{6-k} Q_{hx}^k  &+&  \sum \limits_{k=1}^3 h_k Q_{1z}^4 Q_{1x}^{4-k} Q_{hx}^k \nonumber
\eea

Values of the main coefficients are:

\begin{table}[H]
\centering
\caption{The values of the coefficients of the polynomial
for energy surfaces of Raman and IR modes obtained in \cite{SPhononics}, \cite{SSubediFerro}. The units of a $Q^m Q^p Q^n$ term are meV $A^{-(m+p+n)}$  }
\label{Table1}
\begin{tabular}{l*{6}{c}r}
Coefficient           & Term      & KTaO$_3$ & Coefficient & Term     & LCO  \\
\hline
$\Omega_h^2 $    & $ Q_{hx}^2$        & 1043.77   &    $\Omega_0^2 $  &   $Q_{IR}^2$      &  1462.3      \\
$\Omega_{1z}^2$        &  $ Q_{1z}^2 $   & 27.06       &   $\Omega_1^2 $ &  $Q_R^2$      &   103.55   \\
$\Omega_{1x}^2$        &  $Q_{1x}^2  $      & 27.06   &  $g$     & $ - Q_{R}^2 Q_{IR}^2$/2 & 46.98  \\
  $a_4$                       &     $ Q_{1z}^4  $    & 47.55  &   $ a_4 $         & $ Q_{R}^4 $  &  8.36 \\
$c_4$                        &   $   Q_{hx}^4 $                    & 63.17   &    $ c_4$   &     $ Q_{IR}^4 $  &   103.5    &   \\
\end{tabular} 
\end{table}

\subsection{Strong resonant driving: chaotic dynamics}

\begin{figure}[H]
\includegraphics[width=60mm]{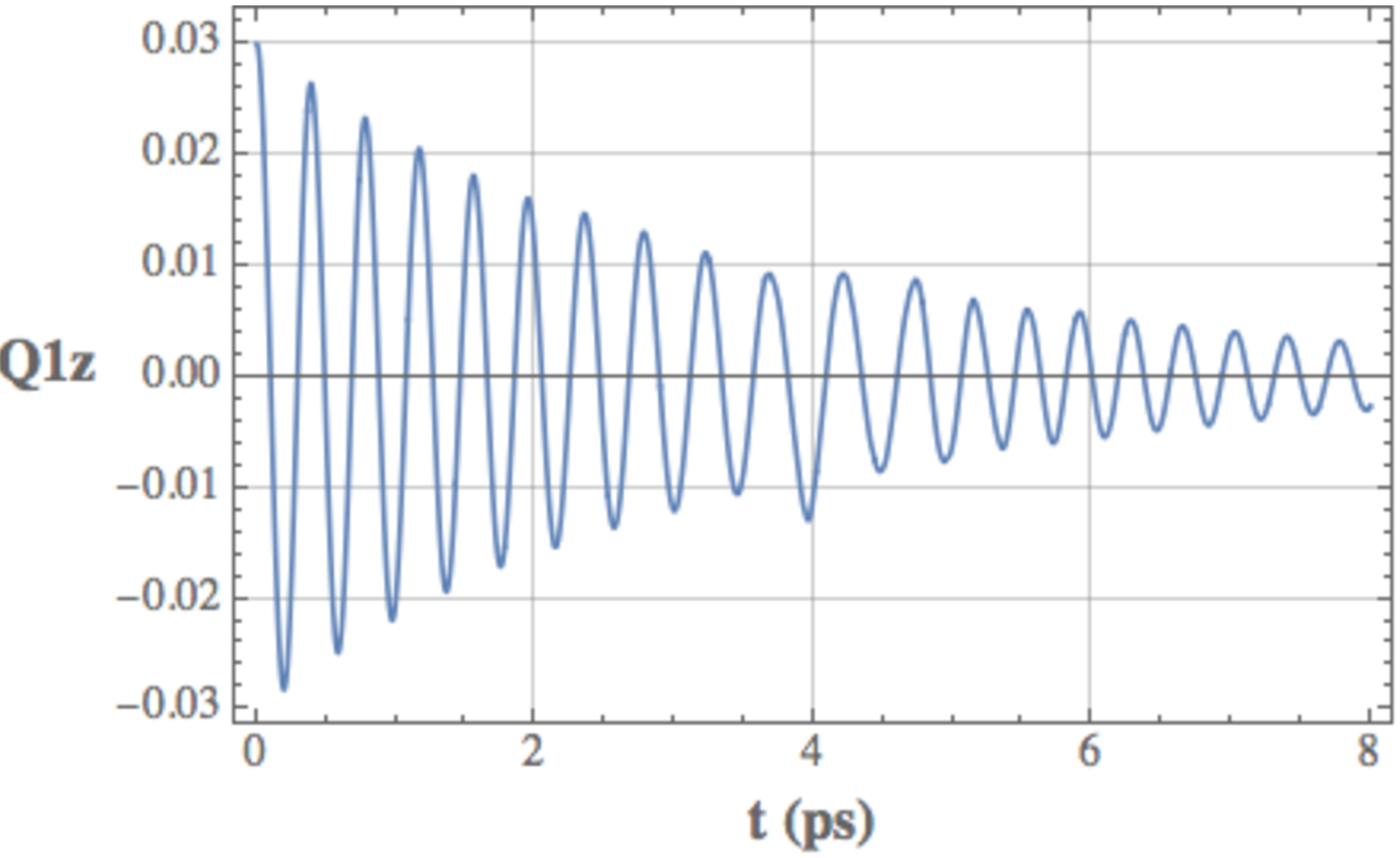}
\includegraphics[width=60mm]{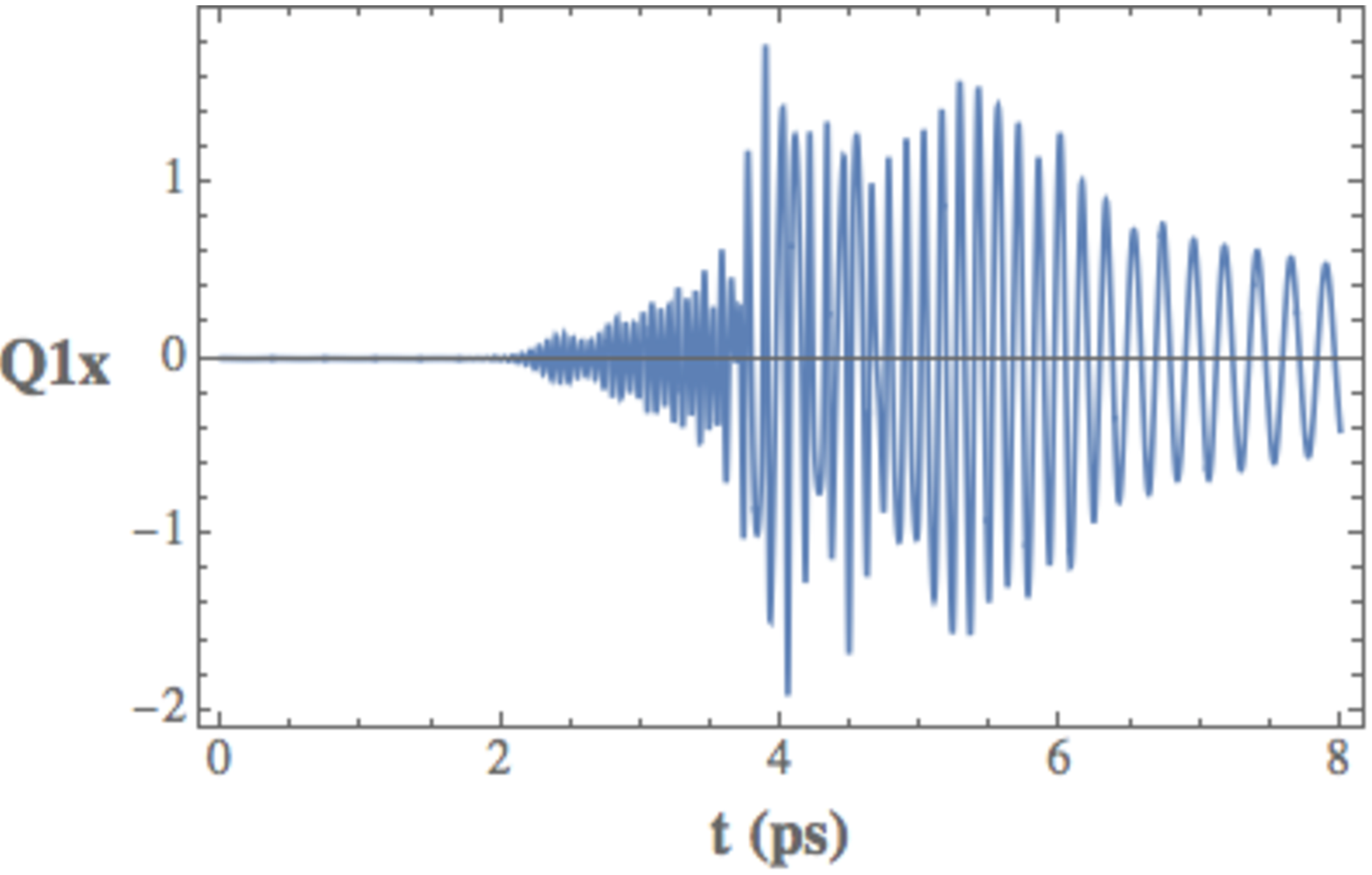}
\includegraphics[width=60mm]{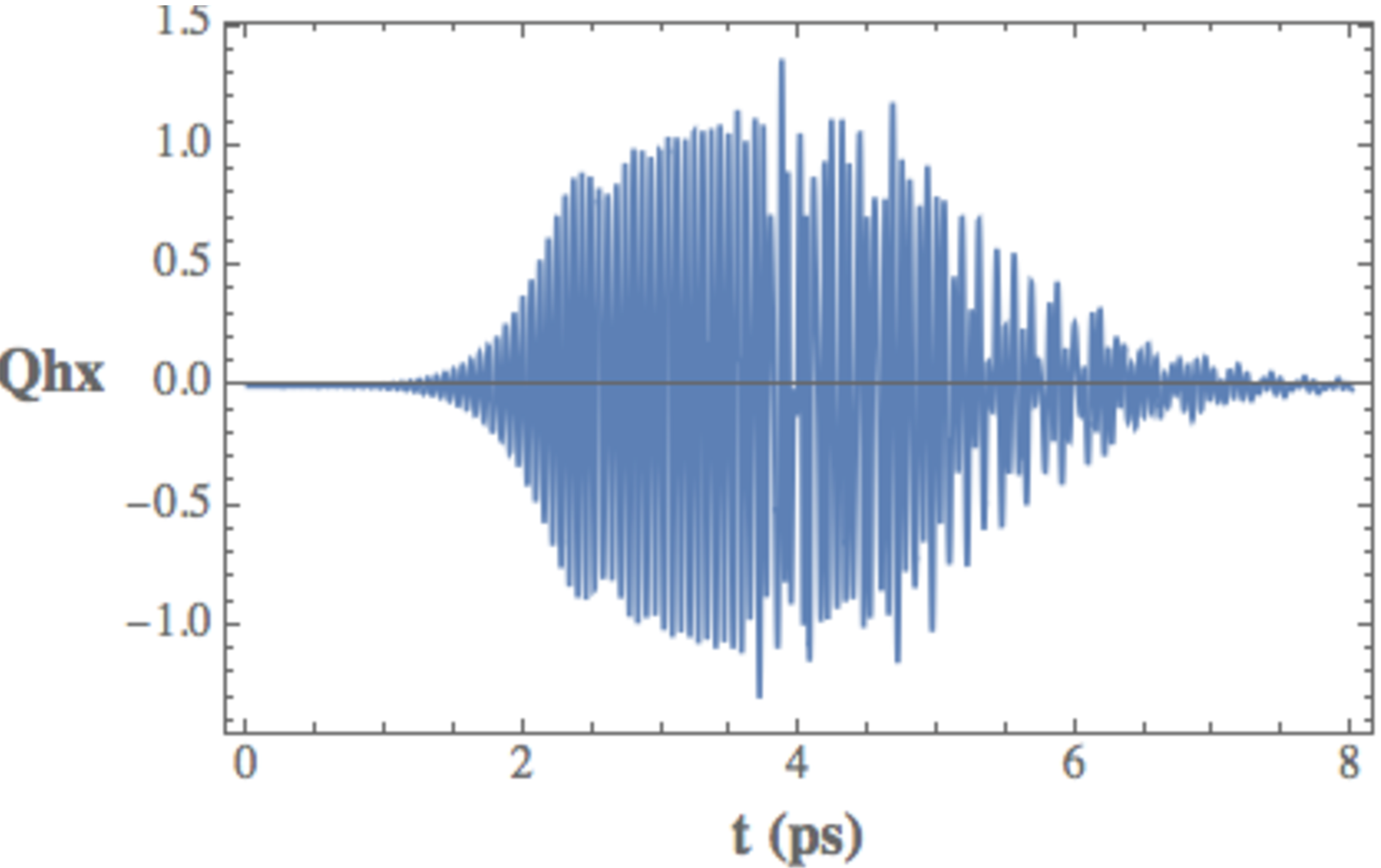}
\caption{ Resonant driving, the same as in Fig.[\ref{E30}], but with a larger driving amplitude.  From left to right:  $Q_{1z}, Q_{1x},Q_{hx}$ modes, and $E(t)$ pulse. $E_0=60$ MV cm$^{-1}$. Amplitude of driving is so large that dynamics become chaotic, nevertheless the mode $Q_{1z}$ is unexcited.   \label{E60}  }
\end{figure}
  
\subsection{Two coupled  modes with periodic driving  } 
 
Although the linear frequency of the $Q_{1x}$ mode is much smaller than that of  $Q_{hx}$,  strong coupling between them causes these modes to oscillate synchronously.
One can therefore introduce an effective system combining these modes together and averaging over the driving frequency. The third mode, $Q_{1z}$ can be neglected until it become excited due to creation of the effective double-well potential.  Consider firstly a conservative system, without dissipation.

The reduced two-mode Hamiltonian is  (we denote $Q_1 \equiv  Q_{1x}, Q_h \equiv  Q_{hx} $)

\bea
H &=& \frac{P_{1}^2}{2} + \frac{P_{h}^2}{2} + \Omega_1^2  \frac{Q_{1}^2}{2}  +  \Omega_h^2  \frac{Q_{h}^2}{2} + a_4 Q_1^4 + c_4 Q_h^4 + \nonumber\\
&+&  t_1 Q_{1}^3 Q_{h} + t_2  Q_{1}^2 Q_{h}^2 + t_3  Q_{1} Q_{h}^3    +  \sum \limits_{k=1}^5 u_k Q_{1}^{6-k} Q_{h}^k +  \nonumber\\
&-& Q_h F_0 \sin \Omega t
 \eea
Moreover, we neglect the $u-$terms in the analytical considerations below, as values of $t-$coefficients are considerably higher.

Introducing radial coordinates ($I_h,\phi_h,I_1,\phi_1$)

\bea
Q_{hx} &=& \sqrt{ \frac{2I_h}{\Omega_h}} \sin \phi_h, \quad P_h =  \sqrt{ 2I_h \Omega_h } \cos \phi_h,  \nonumber\\
Q_ {1x}&=& \sqrt{ \frac{2I_1}{\Omega_1}} \sin \phi_1, \quad P_1 =  \sqrt{ 2I_1 \Omega_1 } \cos \phi_1,
\eea
we then switch to rotating phases $\gamma_h - \Omega t$, $\gamma_1 - \Omega t$ by means of a generating function 
\be W = \rho_h(\phi_h-\Phi) +  \rho_1(\phi_1-\Phi), \Phi \equiv \Omega t \ee

Averaging the resulting Hamiltonian over the explicit time dependence, we get
as an effective  reduced two-mode Hamiltonian 

\bea
H &=& \rho_h (\Omega_h-\Omega)  + \rho_1 (\Omega_1-\Omega) + \frac{3}{2} c_4 \frac{\rho_h^2}{\Omega_h^2} + \frac{3}{2} a_4 \frac{\rho_1^2}{\Omega_1^2}  \nonumber\\
&+& \frac{F_0}{2} \sqrt{ \frac{2\rho_h}{\Omega_h} } \sin \gamma_h   
+  \frac{3}{2} t_1 \Bigl( \frac{\rho_1}{\Omega_1}\Bigr)^{3/2}  \Bigl( \frac{\rho_h}{\Omega_h} \Bigr)^{1/2} \cos (\gamma_1-\gamma_h)    \nonumber\\ 
&+&   \frac{3}{2} t_3 \Bigl( \frac{\rho_1}{\Omega_1} \Bigr)^{1/2}  \Bigl( \frac{\rho_h}{\Omega_h} \Bigr)^{3/2} \cos (\gamma_1-\gamma_h) \\ 
&+&  \frac{1}{2} t_2 \Bigl( \frac{\rho_1}{\Omega_1} \Bigr)  \Bigl( \frac{\rho_h}{\Omega_h} \Bigr)\Bigl(2+  \cos 2(\gamma_1-\gamma_h) \Bigr) + \mbox{u-terms } +... \nonumber
  \eea

We then return to cartesian coordinates via  
$x_k=\sqrt{ \frac{2 \rho_k}{\Omega_k}} \cos \phi_k $, $y_k=\sqrt{ 2 \rho_k \Omega_k} \sin \phi_k $, 
and search for equilibria of the resulted two-mode system.

We get two coupled algebraic equations for coordinates of the equilibrium:

\bea
\frac{3}{2}c_4 X_h^3  + X_h \Bigl( \Omega_h^2(1-x) + \frac{3}{4} t_2 X_1^2 \Bigr) &+& \frac{9}{8} t_3 X_1 X_h^2  \label{Coupled}\\ 
- \frac{3}{8} t_1 X_1^2 &=&\frac{F_0}{2}, \nonumber\\
X_1 \Bigl( \Omega_1^2(1 -x  \frac{\Omega_h}{\Omega_1}) + \frac{3}{4} t_2 X_h^2 \Bigr) &+& \frac{9}{8} t_1 X_1^2 X_h  \nonumber\\
+ \frac{3}{2} a_4 X_1^3 -\frac{3}{8} t_3X_h^3 &=& 0 
\eea

\begin{figure}
\includegraphics[width=80mm]{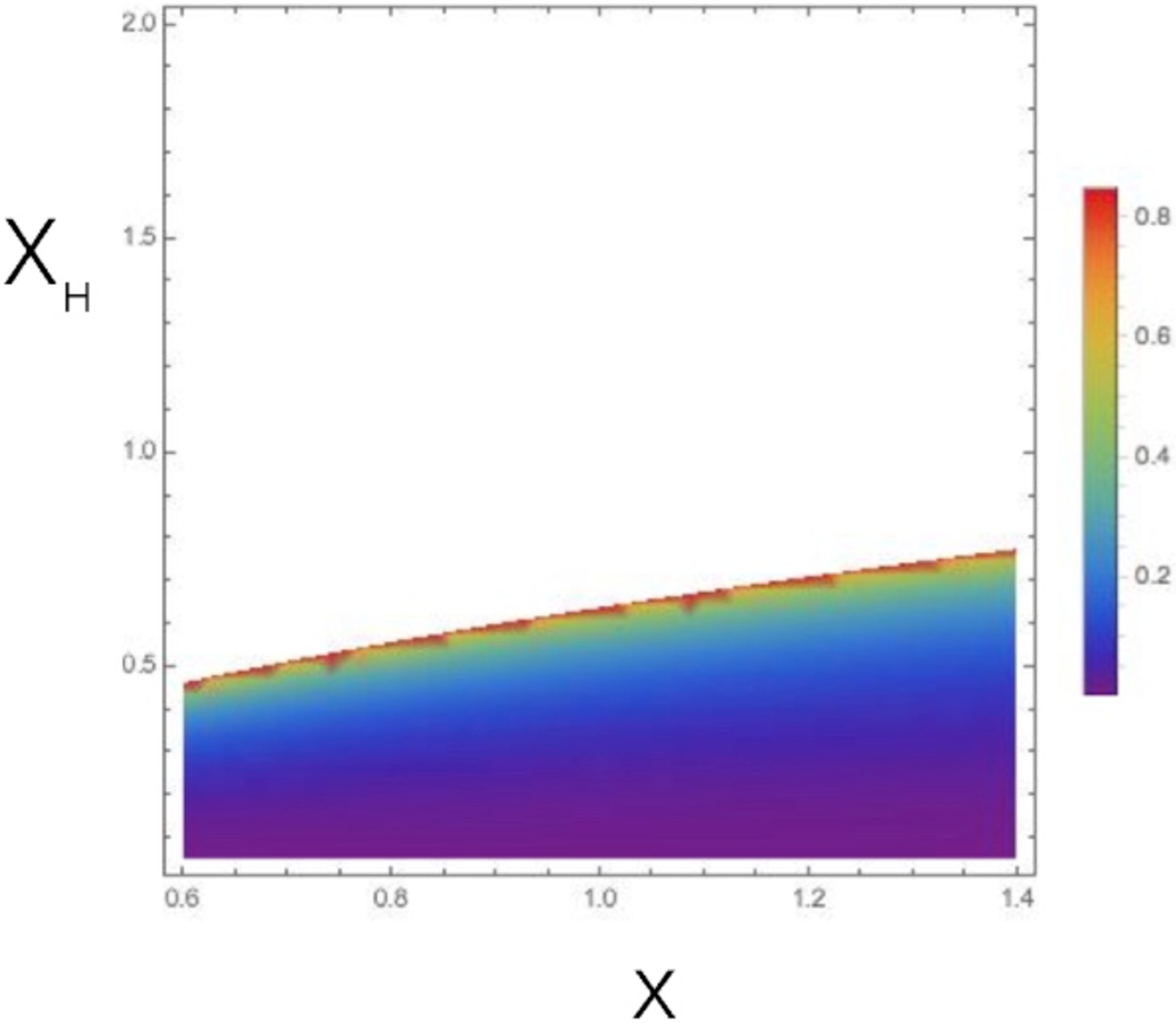}
\caption{  Steady states of the mode $X_h$ of the driven reduced two-mode system (\ref{Coupled}) without dissipation  as a function of frequency $x \equiv \Omega/\Omega_{hx}$ at several values of the driving force amplitude $E_0$.
 From bottom to top: $E_0=5,10,15,20,25,30$ MeV  cm$^{-1}$. (b) Steady values of amplitude of the mode $Q_{1x}$ (i.e., $X_1$) as a function of $X_h, x$.  White areas correspond to highest values of $X_1$    \label{nondissipative}}
\end{figure}

As amplitude of $X_h$ grows beyond certain critical value,  deviation of $X_1$ from 0 becomes considerable. Then, dynamics of the system becomes complicated and the first mode stops to absorb the energy from the drive.
Note that the steady state value of $X_h$ in  Eq. \ref{Coupled}  is a function of $F_0$ and depends on $X_1$, while $X_1$ is determined
from the second equation which do not depends on $F_0$. So we could depict a value of $X_1$ as a density plot on the plane $(\Omega,X_h)$: see Fig. \ref{nondissipative}b.
It is clearly seen that as one tries to increase steady amplitude  $X_h$, above certain curve on $(\Omega,X_h)$ plane the mode $Q_{1x}$ becomes excited.
Importantly, detuning from the exact resonance to higher frequencies allows to reach higher values of $X_h$ without considerable excitation of  the  $Q_{1x}$ mode.  This important qualitative result remains valid in the full system. When we take into account dissipation and the remaining nonlinear coupling terms, the critical curve goes higher, and it become possible to reach sufficiently high values of $X_h$ (again, shifting from the exact resonant to the higher frequencies allows higher values of $X_h$).

\end{document}